\documentclass[12pt]{article}

\usepackage{amsmath}
\usepackage{amsthm}
\usepackage{amsfonts}          
\usepackage{amssymb}           
\usepackage{graphicx}
\usepackage{color}
\usepackage{bbold}

\usepackage[isu,bf]{caption2}
\newlength{\xtrawidth}
\newlength{\xtraheight}
\usepackage[all]{xy}           

\newcommand{\Span}{\mathop{\rm span}\nolimits}
\newcommand{\eqdef}{=}
\newcommand{\Q}{\ensuremath{{\mathbb{Q}}}}
\newcommand{\R}{\ensuremath{{\mathbb{R}}}}
\newcommand{\C}{\ensuremath{{\mathbb{C}}}}

\newcommand{\Z}{\mathbb{Z}}
\newcommand{\CP}{\ensuremath{\mathop{\null {\mathbb{P}}}}\nolimits}

\newcommand{\Tr}{\mathop{\rm Tr}\nolimits}

\newcommand{\Mat}{\mathop{\rm Mat}\nolimits}

\newcommand{\Aut}{{\mathop{\rm Aut}\nolimits}}
\newcommand{\contr}{{\mathop{\rm contr}\nolimits}}
\newcommand{\Osheaf}{{\mathop{\null\mathcal O}\nolimits}}
\newcommand{\Msheaf}{{\mathop{\null\mathfrak M}\nolimits}}

\newcommand{\CY}{Calabi-Yau}
\newcommand{\CYm}{\CY{} manifold}

\newcommand{\YM}{Yang-Mills}
\newcommand{\MW}{Mordell-Weil}
\newcommand{\MWgrp}{\MW{} group}
\newcommand{\MWlat}{\MW{} lattice}

\newcommand{\Hol}{\mathrm{Hol}}

\newcommand{\Pic}{\mathop{\mathrm{Pic}}\nolimits}

\newcommand{\Ncal}{\mathcal{N}}

\newcommand{\Tor}{{\mathop{\rm Tor}\nolimits}}

\newcommand{\Id}{{\mathop{\rm id}\nolimits}}
\newcommand{\rank}{{\mathop{\rm rank}\nolimits}}

\newcommand{\Spin}{{\mathop{\text{\textit{Spin}}}\nolimits}}
\newcommand{\sofrak}{{\mathop{\mathfrak{so}}\nolimits}}
\newcommand{\sufrak}{{\mathop{\mathfrak{su}}\nolimits}}
\newcommand{\hfrak}{{\mathop{\mathfrak{h}}\nolimits}}
\newcommand{\soTenC}{{\mathop{\sofrak(10)_\C}\nolimits}}
\newcommand{\Rep}[1]{\ensuremath{\mathbf{\underline{#1}}}}

\newcommand{\ptset}{\ensuremath{\{\text{pt.}\}}}

\newcommand{\Xt}{\ensuremath{\widetilde{X}}}

\newcommand{\textdef}[1]{{#1}}

\newenvironment{descriptionlist}{%
\begin{list}%
{}%
{}}%
{\end{list}}

{\everymath{\displaystyle\everymath{}}\array}%
{\endarray}

\begin{document}

\begin{titlepage}
  \begin{flushright}
    hep-th/0410055
    \\
    UPR-1089-T
  \end{flushright}
  \vspace*{\stretch{3}}
  \begin{center}
     \Huge Elliptic Calabi-Yau Threefolds with $\Z_3\times
       \Z_3$ Wilson Lines
  \end{center}
  \vspace*{\stretch{2}}
  \begin{center}
    \begin{minipage}{\textwidth}
      \begin{center}
        \large 
        Volker Braun${}^{1}$, Burt A. Ovrut${}^{1}$, 
        Tony Pantev${}^{1}$,
        \\
        and Ren\'e Reinbacher${}^{2}$
      \end{center}
    \end{minipage}
  \end{center}
  \vspace*{1mm}
  \begin{center}
    \begin{minipage}{\textwidth}
      \begin{center}
        ${}^{1}$
        David Rittenhouse Laboratory, University of Pennsylvania\\
        209 S. 33rd Street, Philadelphia, PA 19104, USA
      \end{center}
    \end{minipage}
    \\[5mm]
    \begin{minipage}{\textwidth}
      \begin{center}
        ${}^{2}$ 
        Department of Physics and Astronomy, Rutgers University\\
        136 Frelinghuysen Road, Piscataway, NJ 08854-8019, USA
      \end{center}
    \end{minipage}
  \end{center}
  \vspace*{\stretch{1}}
  \begin{abstract}
    \normalsize A torus fibered Calabi-Yau threefold with first
    homotopy group $\Z_3\times\Z_3$ is constructed as a free quotient
    of a fiber product of two $dP_9$ surfaces. Calabi-Yau threefolds
    of this type admit $\Z_3\times\Z_3$ Wilson lines. In conjunction
    with $SU(4)$ holomorphic vector bundles, such vacua lead to
    anomaly free, three generation models of particle physics with a
    right handed neutrino and a $U(1)_{B-L}$ gauge factor, in addition
    to the $SU(3)_C\times SU(2)_L\times U(1)_Y$ standard model gauge
    group. This factor helps to naturally suppress nucleon decay.  The
    moduli space and Dolbeault cohomology of the threefold is also
    discussed.
  \end{abstract}
  \vspace*{\stretch{5}}
  \begin{minipage}{\textwidth}
    \underline{\hspace{5cm}}
    \centering
    \\
    Email: 
    \texttt{vbraun@physics.upenn.edu},
    \texttt{ovrut@elcapitan.hep.upenn.edu},
    \\ \indent
    \texttt{tpantev@math.upenn.edu},
    \texttt{rreinb@physics.rutgers.edu}.
  \end{minipage}
\end{titlepage}

\newpage



\section{Introduction}
\label{sec:intro}

A long standing question has been whether or not there exist, within
the context of both weakly~\cite{Gross:1985fr, Gross:1985rr} and
strongly coupled~\cite{Horava:1995qa, Horava:1996ma} $E_8\times E_8$
heterotic string theory, vacua that accurately describe low energy
particle physics. One conceptually simple approach has been to
compactify heterotic string theory on a smooth \CY{} threefold $X$
which admits a gauge connection satisfying the hermitian \YM{}
equations.

It was shown in~\cite{MR86h:58038, MR88i:58154} that this latter
requirement is equivalent to demonstrating the existence of a stable,
holomorphic vector bundle $V$ on $X$. In early work, $V$ was chosen to
be the tangent bundle of $X$; that is, $V=TX$. This so-called standard
embedding is reviewed in~\cite{Green:1987sp, Green:1987mn}. While
interesting, this choice is extremely restrictive, fixing $V$ to be
one out of an enormous number of possible vector bundles. Note, for
example, that since $X$ is a \CY{} threefold, the structure group of
$TX$ must be $SU(3)$. Thus, the low energy gauge group associated with
$V=TX$, ignoring possible Wilson lines, is $E_6$, the commutant of
$SU(3)$ in $E_8$. Although $E_6$ is a possible grand unified group,
other choices, such as $SU(5)$ or $\Spin(10)$, would be physically
more interesting.

It is important, therefore, to construct stable holomorphic vector
bundles $V\not=TX$. Recently, it was shown how to obtain such bundles
over simply connected \CY{} threefolds that are elliptically
fibered~\cite{Friedman:1997ih, Donagi:1998xe}. This work was extended,
within the context of heterotic M-Theory~\cite{Lukas:1997fg,
  Lukas:1998ew, Lukas:1998hk, Lukas:1998tt, Lukas:1998yy},
in~\cite{Donagi:1998xe, Donagi:1999gc}. These new vector bundles have
arbitrary structure groups, such as $SU(5)$ and $SU(4)$. These lead to
a wide range of unified gauge groups, including $SU(5)$ and
$\Spin(10)$. Many of the physical properties of these generalized
vacua have been studied, such as the moduli space of the associated
$M5$-branes~\cite{Donagi:1999jp}, small instanton phase
transitions~\cite{Ovrut:2000qi}, fluxes~\cite{Krause:2000gp,
  Curio:2000dw, Curio:2001qi, Curio:2003ur, Cardoso:2003af,
  Cardoso:2002hd}, supersymmetry breaking~\cite{Lukas:1999kt}, the
moduli space of the vector bundle $V$~\cite{Buchbinder:2002pr,
  Buchbinder:2002ji}, and non-perturbative
superpotentials~\cite{Buchbinder:2002pr, Buchbinder:2002ic,
  Lima:2001nh, Lima:2001jc}. Recently, it was shown how to compute the
sheaf cohomology of $V$, as well as that of its tensor products. This
determines the complete particle spectrum~\cite{Donagi:2004ia,
  Donagi:2004qk}. These vacua also underly the theory of brane
universes~\cite{Lukas:1997fg, Lukas:1998ew, Lukas:1998hk,
  Lukas:1998tt, Lukas:1998yy} and ekpyrotic
cosmology~\cite{Khoury:2001zk, Khoury:2001bz, Donagi:2001fs,
  Khoury:2001wf} in strongly coupled heterotic strings.

However, heterotic vacua that have the standard model gauge group
$SU(3)_C \times SU(2)_L \times U(1)_Y$ as a factor need additional
ingredients.  First of all, the \CY{} threefold must have a
non-trivial first homotopy group $\pi_1(X)\not=1$. Only such manifolds
will admit Wilson lines~\cite{Witten:1985xc, Sen:1985eb, Evans:1985vb,
  Breit:1985ud, Breit:1985ns}. One must then construct stable,
holomorphic vector bundles on these spaces which, in conjunction with
Wilson lines, will reduce the gauge group so as to include
$SU(3)_C\times SU(2)_L\times U(1)_Y$ as a factor. Torus fibered \CY{}
threefolds with $\pi_1(X)=\Z_2$ were considered
in~\cite{Donagi:2000fw, Donagi:2000zs, Donagi:2000zf, Donagi:1999ez}.
These manifolds admit stable, holomorphic vector bundles with
structure group $SU(5)$. Together with the $\Z_2$ Wilson lines, such
vacua produce anomaly free theories with three generations, and gauge
group exactly $SU(3)_C \times SU(2)_L \times U(1)_Y$. In recent work,
the complete cohomology ring and, hence, the spectrum of such vacua
was computed. These models were found to have at least one pair of
Higgs doublets, but contain exotic supermultiplets as well. This work
was generalized in~\cite{Donagi:2003tb, Ovrut:2003zj, Ovrut:2002jk} to
torus fibered \CY{} threefolds with $\pi_1(X)=\Z_2\times\Z_2$. These
were shown to admit stable, holomorphic vector bundles with structure
group $SU(4)$.

The recent discovery that neutrinos have a nonzero
mass~\cite{Fukuda:1998mi} has made low energy theories based on
spontaneously broken $\Spin(10)$ very attractive. There are several
reasons for this. First, such theories naturally contain a right
handed neutrino. Second, each family of quarks and leptons is unified
within a single $\Rep{16}$, the spin representation of $\Spin(10)$. A
third compelling reason is that these theories can contain an extra
$U(1)_{B-L}$ gauge factor, which greatly helps to suppress the nucleon
decay rate. For all these reasons, it would seem desirable to
construct heterotic string vacua of this type. In this paper, we are
taking the first steps in this direction. We begin by demonstrating
that the simplest Wilson line that breaks $\Spin(10)$ to the standard
model with an extra $U(1)_{B-L}$ requires a $\Z_3\times\Z_3$
fundamental group. We then systematically construct torus fibered
\CY{} threefolds with this fundamental group. Several properties of
these geometries will also be computed, including the Dolbeault
cohomology groups. Stable, holomorphic vector bundles on these \CY{}
threefolds will be presented elsewhere~\cite{toappear}

The paper is organized as follows. First, in
Section~\ref{sec:symbreaking}, we review how a $\Z_3\times \Z_3$
Wilson line can break a $\Spin(10)$ gauge group down to the standard
model plus a $U(1)_{B-L}$ factor. The general theory and the
intermediate $\Spin(10)$ gauge group is introduced in
Subsection~\ref{sec:symgen}.  Then, in~\ref{sec:break1}
and~\ref{sec:break2} we successively consider the effect of two
distinct $\Z_3$ Wilson lines. These two effects are combined in
Subsection~\ref{sec:symbreakcombined}, and the correct low energy
gauge group is obtained.

In Section~\ref{sec:CYintro}, we analyze how one can implement this
symmetry breaking within heterotic string theory. This imposes various
restrictions on the compactification manifold. We will choose the
\CY{} threefold to be the fiber product of two $dP_9$ surfaces.

It follows that one must study $dP_9$ surfaces whose automorphism
group contains a $\Z_3\times\Z_3$ subgroup. This will be done in
Section~\ref{sec:surfaces}. First, the general topology of $dP_9$
surfaces is reviewed in Subsection~\ref{sec:surfacesGaction}. We then
study the general form of $\Z_3\times\Z_3$ actions on these surfaces
using the \MWgrp{} and its bilinear form. These are introduced in
Subsections~\ref{sec:MW} and~\ref{sec:MWlat} and then applied to the
$dP_9$ surfaces in~\ref{sec:singularfib}. There are only a small
number of allowed \MWgrp{}s, and we use these to list all $dP_9$
surfaces having the desired properties. Their moduli are also
discussed.  Finally, in Subsection~\ref{sec:fixedpoints} we
investigate which surfaces yield fiber products with free
$\Z_3\times\Z_3$ actions. We conclude that only a single $1$-parameter
family is appropriate.

This $1$-parameter family of $dP_9$ surfaces is constructed in
Section~\ref{sec:explicit}. To begin with, the Weierstrass model of an
elliptic surface is reviewed in Subsection~\ref{sec:Wgeneral}. Then,
we write down the Weierstrass equation for the desired $dP_9$ surface
in Subsection~\ref{sec:weierstrass}, thus establishing its existence.
We check explicitly that it has the correct singular fibers and the
desired group action. For completeness, the surface is also described
as a pencil of cubics in Subsection~\ref{sec:pencil}.

It remains to study the homology of this $dP_9$ surface and the
induced action on the homology of the $\Z_3\times \Z_3$ automorphism.
This is the aim of Section~\ref{sec:dP9homology}. First, in
Subsection~\ref{sec:MWweierstrass}, we investigate how the singular
fibers in the $dP_9$ surface intersect the sections of the elliptic
fibration. Using this information, the $\Z_3\times\Z_3$ action on the
homology of the $dP_9$ surface is determined in
Subsection~\ref{sec:surfaceHaction}.
In~\ref{sec:surfaceinvarianthomology}, the invariant part of the
homology is finally computed.

Having studied these $dP_9$ surfaces in detail, we construct in
Section~\ref{sec:CY} the desired \CY{} threefold. The fiber product of
two such $dP_9$ surfaces is a \CYm{} $\Xt$. Furthermore, if one makes
the correct identifications in the fiber product, as discussed in
Subsection~\ref{sec:CYfiberproduct}, then there exists a free
$G=\Z_3\times\Z_3$ action on $\Xt$. The homology of $\Xt$ and the $G$
action on it are discussed in Subsection~\ref{sec:homologyCYX}.
In~\ref{sec:homologyCYXt}, we form the quotient $X=\Xt/G$.  This is a
\CY{} threefold with fundamental group $\pi_1(X)=\Z_3\times\Z_3$, as
desired.  Its Hodge diamond is then computed and its moduli are
discussed.

\section{Breaking to the Standard Model Gauge Group}
\label{sec:symbreaking}

\subsection{Symmetry Breaking Generalities}
\label{sec:symgen}

In this paper, we consider \CY{} threefolds $X$ with non-vanishing
first homotopy group\footnote{More precisely, we only consider
  ``proper'' \CY{} threefolds, that is, compact with the full $SU(3)$
  holonomy.  The fundamental group is then necessarily finite.}
$\pi_1(X)$. It is further assumed that these spaces admit stable,
holomorphic vector bundles $V$ with structure group
\begin{equation}
  \label{eq:SU4E8}
  G_V = SU(4) \subset E_8
  \,.
\end{equation}
Specifying $X$ and $V$ determines a vacuum of $E_8\times E_8$
heterotic string theory. This vacuum has, at low energies, $\Ncal=1$
supersymmetry in $\R^4$ with gauge symmetry
\begin{equation}
  \label{eq:Spin10}
  H_{SU(4)} = \Spin(10) 
\end{equation}
in the observable sector. The low energy gauge group $H_{SU(4)}$ is
the commutant of $SU(4)$ in $E_8$. 

Since $\pi_1(X)$ is non-trivial, one can consider, in addition to $V$,
a vector bundle $W$ on $X$ with a discrete structure group
\begin{equation}
  G_W \simeq \pi_1(X) \subset \Spin(10) \subset E_8
  \,.
\end{equation}
$W$ admits a unique flat connection with holonomy group
\begin{equation}
  \label{eq:pi1HolF}
  {1} \not= \Hol(W) = \pi_1(X)
  \,.
\end{equation}
We can then define another vacuum of the $E_8\times E_8$ heterotic
string by considering the vector bundle $V\oplus W$ on $X$. This
preserves $\Ncal=1$ supersymmetry, but the low energy gauge group is
reduced to
\begin{equation}
  \label{eq:brokenH}
  H_{SU(4)\times \Hol(W)} 
  \subset
  \Spin(10)
  \,.
\end{equation}

We will assume that $\Hol(W)$ is Abelian in the following. Then
\begin{equation}
  \label{eq:rkH}
  \rank~ H_{SU(4)\times \Hol(W)} =
  \rank~ \Spin(10) = 5
  \,.
\end{equation}
Any realistic string vacuum must incorporate the standard model gauge
group $SU(3)_C \times SU(2)_L \times U(1)_Y$ at low energies, and this
group has rank $4$. It follows from eq.~\eqref{eq:rkH} that,
minimally, $H_{SU(4)\times \Hol(F)}$ must, in addition to the standard
model gauge group, have an extra $U(1)$ gauge factor. It can be shown
that a gauged $U(1)_{B-L}$ group, broken at a scale not much larger
than the electroweak scale, helps to suppress rapid nucleon decay.
Hence, an additional $U(1)_{B-L}$ is a phenomenologically desirable
component in the low energy gauge group. Therefore, we would like to choose
$W$ so that
\begin{equation}
  \label{eq:brokenHsm}
  H_{SU(4)\times \Hol(W)} =  
  SU(3)_C \times SU(2)_L \times U(1)_Y \times U(1)_{B-L}
  \,.
\end{equation}
In this section, we will show that this can be arranged provided that
\begin{equation}
  \label{eq:holFZ3Z3}
  \Hol(W)\simeq \Z_3\times \Z_3
  \,.
\end{equation}
Other choices of $\Hol(W)$ can also lead to eq.~\eqref{eq:brokenHsm}
but, within the context of our construction, $\Z_3\times \Z_3$ is the
smallest group.

We begin by recalling some basic facts about the complexified Lie
algebra $\soTenC$ of $\Spin(10)$. First, it is a simple Lie algebra
with
\begin{equation}
  \label{eq:soTenCdim}
  \dim \soTenC = 45
  ,\quad
  \rank~ \soTenC = 5
  \,.
\end{equation}
Up to conjugation, there exists a unique maximal Abelian subalgebra
$\hfrak$, the Cartan subalgebra, with
\begin{equation}
  \label{eq:hdim}
  \dim \hfrak = 5
  \,.
\end{equation}
Since $[\hfrak,\soTenC] \subset \soTenC$, the Lie algebra is a module
carrying a (fully reducible) linear representation of $\hfrak$. As
$\hfrak$ is Abelian, every irreducible subspace is one-dimensional.
This leads to the Cartan-Weyl decomposition of $\soTenC$ given by
\begin{equation}
  \label{eq:soTenCCartanWeyl}
  \soTenC = \hfrak \oplus 
  \sum_\alpha \C e_\alpha
  \,.
\end{equation}
Each $\C e_\alpha$ is a one-dimensional $\hfrak$ module which we can
label by some
\begin{equation}
  \label{eq:alphahfrakast}
  \alpha \in \hfrak^\ast
  \,,
\end{equation}
where $\hfrak^\ast$ denotes the dual space to $\hfrak$. Each $\alpha$
occurring in eq.~\eqref{eq:soTenCCartanWeyl} is called a root of
$\soTenC$, and it follows from eq.~\eqref{eq:soTenCdim} that $\soTenC$
has forty roots. Let $\Phi$ be this set of roots. They span
$\hfrak^\ast$, but cannot all be linearly independent. Hence, one can
choose a basis $\Delta\subset \Phi$. Actually, $\Delta$ can be chosen
such that each root $\alpha\in \Phi$ can be written as a $\Z$-linear
combination of the roots in $\Delta$ with the integral coefficients
either all $\geq 0$ or all $\leq 0$. The elements of $\Delta$ are
called the simple roots. Since
$\dim \hfrak^\ast=5$, there are five simple roots of
$\soTenC$, which we denote by
\begin{equation}
  \Delta \eqdef \Big\{ \alpha^i ~\Big|~ i=1,\dots,5\Big\}
  \,.
\end{equation}
To determine the roots explicitly, it is helpful to consider the
linear subspace
\begin{equation}
  \hfrak^\ast_\R = \Span_\R \Delta
  \,.
\end{equation}
Clearly
\begin{equation}
  \dim_\R \hfrak^\ast_\R = \dim_\C \hfrak^\ast = 5
  ,\quad
  \Phi \subset \hfrak^\ast_R \subset \hfrak^\ast
  \,.
\end{equation}
The Killing form on $\soTenC$ is defined by
\begin{equation}
  \label{eq:Killing}
  (x,y) \eqdef \frac{1}{\lambda} 
  \Tr\Big( \mathrm{ad}(x) \mathrm{ad}(y) \Big)
  \qquad \forall x,y\in \soTenC
  \,,
\end{equation}
where the value of the normalization constant $\lambda$ will be
specified below. Since $\soTenC$ is simple, the Killing form is
non-degenerate. This allows us to associate with any $x^\ast \in
\hfrak^\ast$ a unique element $x\in \hfrak$ by
\begin{equation}
  x^\ast(y) = (x,y) 
  \qquad \forall y\in\soTenC
  \,.
\end{equation}
One can then define a bilinear form $\hfrak^\ast \times \hfrak^\ast
\to \C$ via
\begin{equation}
  \langle x^\ast, y^\ast\rangle \eqdef (x,y)
  \,.
\end{equation}
This form and its restriction to $\hfrak^\ast_\R$ is again
non-degenerate. Furthermore, when restricted to $\hfrak^\ast_\R$, it
is positive definite.

\begin{figure}[htbp]
  \centering \input{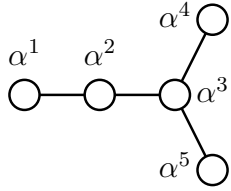}
  \caption{The $\Spin(10)$ Dynkin diagram.}
  \label{fig:Spin10Dynkin}  
\end{figure}

The roots of $\soTenC$ are the following. Consider the Euclidean space
$\hfrak^\ast_\R \simeq \R^5$ and let $e^i$, $i=1,\dots,5$ be an
orthonormal basis. Then the five simple roots are
\begin{equation}
\label{eq:simpleroots}
\begin{split}
  \alpha^i &= e^i - e^{i+1}, \quad i=1,\dots,4
  \\
  \alpha^5 &= e^4+e^5 \,.
\end{split}
\end{equation}
The remaining thirty-five roots can be obtained by acting on
eq.~\eqref{eq:simpleroots} with Weyl reflections and using the fact
that if $\alpha$ is a root, then so is $-\alpha$. There is no need to
construct them explicitly. The associated Dynkin diagram is shown in
Figure~\ref{fig:Spin10Dynkin}.

A Chevalley basis of $\soTenC$ consists of five elements $h_i$,
$i=1,\dots,5$ which span $\hfrak$ along with the forty elements
$e_\alpha$, $\alpha \in \Phi-\Delta$. In addition,
these satisfy five commutator relations of which we will use the
following two:
\begin{subequations}
  \begin{align}
    \big[h_i,~ h_j\big] &= 0
    \\
    \big[h_i,e_\alpha\big] &= \alpha(h_i) e_\alpha \,.
  \end{align}
\end{subequations}
We then choose the normalization of the Killing form,
eq.~\eqref{eq:Killing}, so that
\begin{equation}
  \big(h_i,h_j\big) = \delta_{ij}
  ,\quad
  \big(e_\alpha,e_\beta\big) = \delta_{\alpha\beta}
\end{equation}
for all $i$, $j$ and $\alpha$, $\beta$. We are free to choose each
$h_i$ to be the dual element in $\hfrak$ of $e^i\in \hfrak^\ast$. That
is
\begin{equation}
  \label{eq:ehdual}
  e^i\big( h_j \big) = \delta^i_j
  ,\quad 
  i,j=1,\dots,5
  \,.
\end{equation}
For the choice of simple roots in eq.~\eqref{eq:simpleroots}, it then
follows that one must take
\begin{equation}
  \lambda = 16
  \,.
\end{equation}
We will use this normalization for the Killing form henceforth.

\subsection{First Wilson Line}
\label{sec:break1}

\begin{figure}[htbp]
  \centering \input{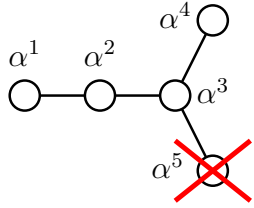}
  \caption{$\Spin(10)$ Dynkin diagram with the root $\alpha^5$ removed.}
  \label{fig:Spin10DynkinX5}
\end{figure}

Let us now find the $\sufrak(2)_\C$ subalgebra associated with the
simple root $\alpha^5$. We denote the generator of its Cartan
subalgebra by $H_{(5)}$. It is clear from the marked Dynkin diagram in
Figure~\ref{fig:Spin10DynkinX5} that the product of $H_{(5)}$ with any
element of the $\sufrak(5)_\C$ subalgebra associated with the simple
roots $\alpha^1,\dots,\alpha^4$ must vanish. Since each of the
$\sufrak(5)_\C$ roots is a linear combination of the four simple roots,
it suffices to show that
\begin{equation}
  \label{eq:HEcommutator}
  \left[ H_{(5)},~ E_{\alpha^i} \right] =0
  ,\quad i=1,\dots,4
  \,.
\end{equation}
Writing
\begin{equation}
  \label{eq:Hexpcoeff}
  H_{(5)} = \sum_{i=1}^5 a^i h_i
\end{equation}
and using eqns.~\eqref{eq:simpleroots} and~\eqref{eq:ehdual}, it
follows that all $a^i$ must be identical and, hence,
\begin{equation}
  H_{(5)} = a (h_1 + h_2 + h_3 + h_4 + h_5)
  \,.
\end{equation}
The coefficient $a$ cannot be determined from the Lie algebra alone.
To find $a$, one must construct the one-parameter Abelian subgroup of
$\Spin(10)$ generated by $H_{(5)}$. We denote this subgroup by
$U(1)_{(5)}$. In every representation of $\Spin(10)$, $U(1)_{(5)}$
must be periodic in its parameter $\theta$. This will be the case if
and only if it is periodic in the $16$-dimensional complex spin
representation $\Rep{16}$. The spin representation of the generators
$h_i$, $i=1,\dots,5$ can be constructed by standard
methods~\cite{MR93a:20069}. We find that
\begin{equation}
  \left[ H_{(5)} \right]_\Rep{16} = \frac{a}{2} 
  \left[ Y_{(5)} \right]_\Rep{16}
  \,,
\end{equation}
where
\begin{equation}
  \left[Y_{(5)}\right]_\Rep{16} = 
  \left(
    \begin{array}{c|c|c}
      -\mathbb{1}_{10} &  \multicolumn{2}{c}{}
      \\ \cline{1-2} 
      & 3 \cdot \mathbb{1}_{5}
      \\ \cline{2-3} 
      \multicolumn{1}{c}{} & & -5
    \end{array}
  \right)
\end{equation}
and $\mathbb{1}_m$ stands for the $m\times m$ unit matrix. An element
in the associated $U(1)_{(5)}$ subgroup is then given by
\begin{equation}
  \label{eq:g5theta}
  \left[g_{(5)}(\theta)\right]_\Rep{16} = 
  \left(
    \begin{array}{c|c|c}
      e^{-\frac{i a}{2}\theta}\mathbb{1}_{10} 
      &  \multicolumn{2}{c}{}
      \\ \cline{1-2} {\vbox{\vspace{5mm}}}
      & e^{\frac{3 i a}{2}\theta} \mathbb{1}_{5} 
      \\ \cline{2-3} 
      \multicolumn{1}{c}{} & {\vbox{\vspace{5mm}}} & 
      e^{-\frac{5 i a}{2}\theta}
    \end{array}
  \right)
  \,,
\end{equation}
where $\theta$ runs over some interval. One can, without loss of
generality, assume that $0\leq\theta<2\pi$, in which case $a$ must be
a multiple of $2$. Henceforth, we choose
\begin{equation}
  a = 2
  \,.
\end{equation}

Having determined $H_{(5)}$ and $U(1)_{(5)}$, we want to consider the
finite subgroup of $U(1)_{(5)}$ generated by the element with
$\theta=\frac{2 \pi}{3}$. We denote this cyclic subgroup by
$(\Z_3)_{(5)}$. This is most easily done by choosing a representation
of $\Spin(10)$. Since we have already constructed it, we will take
this to be the spin representation $\Rep{16}$. It follows from
eq.~\eqref{eq:g5theta} that the generator of $(\Z_3)_{(5)}$ is given
by
\begin{equation}
  \label{eq:g5Z3}
  \left[g_{(5)}\left(\frac{2 \pi}{3}\right)\right]_\Rep{16} = 
  \left(
    \begin{array}{c|c|c}
      e^{-\frac{2\pi i}{3}}\mathbb{1}_{10} 
      &  \multicolumn{2}{c}{}
      \\ \cline{1-2}
      &  \mathbb{1}_{5} 
      \\ \cline{2-3} 
      \multicolumn{1}{c}{} & {\vbox{\vspace{5mm}}} & 
      e^{-\frac{4\pi i}{3}}
    \end{array}
  \right)
  \,.
\end{equation}
Since $e^{-\frac{2\pi i}{3}}\not=1\not=e^{-\frac{4\pi i}{3}}$, it
follows that the commutant of $(\Z_3)_{(5)}$ in $\Spin(10)$ is
$SU(5)\times U(1)_{(5)}$. Therefore, if we choose a flat line bundle
$W$ with holonomy group
\begin{equation}
  \Hol(W) = (\Z_3)_{(5)}
  \,,
\end{equation}
then
\begin{equation}
  H_{SU(4)\times (\Z_3)_{(5)}} = 
  SU(5) \times U(1)_{(5)}
  \,.
\end{equation}
With respect to this subgroup, we can reduce the $\Spin(10)$ spin
representation as
\begin{equation}
  \label{eq:SU5reduce}
  \Rep{16} = 
  (\Rep{10},-1) \oplus 
  (\Rep{\overline{5}},3) \oplus 
  (\Rep{1},-5)
  \,.
\end{equation}

\subsection{Second Wilson Line}
\label{sec:break2}

\begin{figure}[htbp]
  \centering \input{Spin10DynkinX3.pstex_t}
  \caption{$\Spin(10)$ Dynkin diagram with the root $\alpha^3$ removed.}
  \label{fig:Spin10DynkinX3}
\end{figure}
Let us now find the Cartan subalgebra of the $\sufrak(2)_\C$ in
$\soTenC$ associated with the root $\alpha^3$. We denote its generator
by $H_{(3)}$. It is clear from the marked Dynkin diagram in
Figure~\ref{fig:Spin10DynkinX3} that the product of $H_{(3)}$ with any
element of the $\sufrak(3)_\C \oplus \sufrak(2)_\C \oplus
\sufrak(2)_\C$ subalgebra associated with the simple roots $\alpha^1$,
$\alpha^2$, $\alpha^4$ and $\alpha^5$ must vanish. That is
\begin{equation}
  \left[ H_{(3)},~ E_{\alpha^i} \right] =0
  \,, \quad i=1,2,4,5
  \,.
\end{equation}
It follows from our choice of basis, eqns.~\eqref{eq:simpleroots}
and~\eqref{eq:ehdual}, that
\begin{equation}
  H_{(3)} = b \left( h_1 + h_2 + h_3 \right)
\end{equation}
for some constant $b$. Denote by $U(1)_{(3)}$ the one-parameter
subgroup of $\Spin(10)$ generated by $H_{(3)}$. We determine the
parameter $b$ again by considering the spin representation $\Rep{16}$
of $\Spin(10)$ and demanding periodicity in the parameter $0\leq
\theta<2\pi$ of the $U(1)_{(3)}$. In a convenient basis
\begin{equation}
  [H_{(3)}]_\Rep{16} = \frac{b}{2} [Y_{(3)}]_\Rep{16}
  \,,
\end{equation}
where
\begin{equation}
  \left[Y_{(3)}\right]_\Rep{16} = 
  \left(
    \begin{array}{c|c|c|c}
      \mathbb{1}_{6} &  \multicolumn{2}{c}{}
      \\ \cline{1-2} 
      & -\mathbb{1}_{6} &  \multicolumn{2}{c}{}
      \\ \cline{2-3} 
      \multicolumn{1}{c}{} & & 3 \cdot \mathbb{1}_{2}
      \\ \cline{3-4} 
      \multicolumn{2}{c}{} & & -3\cdot\mathbb{1}_2
    \end{array}
  \right)
  \,.
\end{equation}
A general element of the associated $U(1)_{(3)}$ subgroup is then
given by
\begin{equation}
  \label{eq:g3theta}
  \left[g_{(3)}(\theta)\right]_\Rep{16} = 
  \left(
    \begin{array}{c|c|c|c}
      e^{\frac{i b}{2}\theta}\mathbb{1}_{6} &  
      \multicolumn{2}{c}{}
      \\ \cline{1-2} 
      & e^{-\frac{i b}{2}\theta}\mathbb{1}_{6} &  
      \multicolumn{2}{c}{{\vbox{\vspace{5mm}}}}
      \\ \cline{2-3} 
      \multicolumn{1}{c}{{\vbox{\vspace{5mm}}}} 
      & & e^{\frac{3i b}{2}\theta} \mathbb{1}_{2}
      \\ \cline{3-4} 
      \multicolumn{2}{c}{{\vbox{\vspace{5mm}}}} 
      & & e^{-\frac{3 i b}{2}\theta}\mathbb{1}_2
    \end{array}
  \right)
  \,.
\end{equation}
Again, it follows that $b$ must be a multiple of $2$, and we will take
\begin{equation}
  b = 2
  \,.
\end{equation}

Having determined $H_{(3)}$ and $U(1)_{(3)}$, we want to consider the
finite subgroup of $U(1)_{(3)}$ generated by the element with
$\theta=\frac{2\pi}{3}$. We denote this cyclic subgroup by
$(\Z_3)_{(3)}$. This is most easily done by choosing a representation
of $\Spin(10)$, which we take to be $\Rep{16}$. It follows from
eq.~\eqref{eq:g3theta} that the generator of $(\Z_3)_{(3)}$ is given
by
\begin{equation}
  \label{eq:g3Z3}
  \left[g_{(3)}\left(\frac{2\pi}{3}\right)\right]_\Rep{16} = 
  \left(
    \begin{array}{c|c|c|c}
      e^{\frac{2\pi i}{3}}\mathbb{1}_{6} &  
      \multicolumn{2}{c}{}
      \\ \cline{1-2} 
      & e^{\frac{4 \pi i}{3}}\mathbb{1}_{6} &  
      \multicolumn{2}{c}{{\vbox{\vspace{5mm}}}}
      \\ \cline{2-3} 
      \multicolumn{1}{c}{} 
      & & \mathbb{1}_{2}
      \\ \cline{3-4} 
      \multicolumn{2}{c}{} 
      & & \mathbb{1}_2
    \end{array}
  \right)
  \,.
\end{equation}
Since $e^{\frac{2 \pi i}{3}}\not= e^{\frac{4\pi i}{3}}\not =1$, the
commutant of $(\Z_3)_{(3)}$ in $\Spin(10)$ is $SU(3)\times SU(2)^2
\times U(1)_{(3)}$. This is so despite the degeneracy of the two
$\mathbb{1}_2$ blocks, as can be verified in several ways. First, note
that the decomposition of $\Rep{16}$ with respect to any other
subgroup of $\Spin(10)$ is inconsistent with eq.~\eqref{eq:g3Z3}. For
example, with respect to $SU(4)\times SU(2)^2$ the $\Rep{16}$
decomposes as $\Rep{16}=(\Rep{4},\Rep{1},\Rep{2})\oplus
(\Rep{\overline{4}},\Rep{2},\Rep{1})$. Second, one can construct the
embedding of the $(\Z_3)_{(3)}$ generator in the $\Rep{10}$
representation of $\Spin(10)$. We find that
 
\begin{equation}
  \label{eq:g3Z3rep10}
  \left[g_{(3)}\left(\frac{2 \pi}{3}\right)\right]_\Rep{10} = 
  \left(
    \begin{array}{c|c|c}
      e^{\frac{4\pi i}{3}}\mathbb{1}_{3} 
      &  \multicolumn{2}{c}{}
      \\ \cline{1-2} {\vbox{\vspace{5mm}}}
      &  e^{\frac{2\pi i}{3}}\mathbb{1}_{3} 
      \\ \cline{2-3} 
      \multicolumn{1}{c}{} &  & 
      \mathbb{1}_{4}
    \end{array}
  \right)
  \,,
\end{equation} 
which clearly commutes only with the $SU(3)\times SU(2)^2 \times
U(1)_{(3)}$ subgroup of $\Spin(10)$. Therefore, if we choose a flat
line bundle with holonomy group
\begin{equation}
  \Hol(W)=(\Z_3)_{(3)}
  \,,
\end{equation}
then
\begin{equation}
  H_{SU(4)\times (\Z_3)_{(3)}} = 
  SU(3) \times \Big( SU(2) \Big)^2 \times U(1)_{(3)}
  \,.
\end{equation}
With respect to this subgroup
\begin{equation}
  \Rep{16} = 
  \Big( \Rep{3}, \Rep{2}, \Rep{1}, 1 \Big) \oplus
  \Big( \Rep{\overline{3}}, \Rep{1}, \Rep{2}, -1 \Big) \oplus
  \Big( \Rep{1}, \Rep{1}, \Rep{2}, 3 \Big) \oplus
  \Big( \Rep{1}, \Rep{2}, \Rep{1}, -3 \Big)
  \,.
\end{equation}
Thus far, we have used a basis which most easily allowed us to find
the commutant subgroup to $(\Z_3)_{(3)}$. This, however, is not the
same basis as was used to write the $(\Z_3)_{(5)}$ generator in
eq.~\eqref{eq:g5Z3}. Changing the basis to the one used
in~\eqref{eq:g5Z3}, we find that
\begin{equation}
  \label{eq:g3Z3basis}
  \left[ g_{(3)}\left( \frac{2 \pi}{3} \right) \right]_\Rep{16} =
  \left( 
    \begin{array}{c|c|c|c|c|c}
      e^{\frac{2\pi i}{3}} \mathbb{1}_6 
      & \multicolumn{1}{c}{} & 
      \\ \cline{1-2}
      \vbox{\vspace{4mm}} & 
      1 &
      \\ \cline{2-3}
      \multicolumn{1}{c}{} & \vbox{\vspace{5mm}} & 
      e^{\frac{4\pi i}{3}} \mathbb{1}_3 
      \\ \cline{1-5}
      \multicolumn{2}{c}{} & \vbox{\vspace{4mm}} & 
      \mathbb{1}_2 &
      \\ \cline{4-5}
      \multicolumn{2}{c}{} & & \vbox{\vspace{5mm}} & 
      e^{\frac{4\pi i}{3}} \mathbb{1}_3
      \\ \cline{4-6}
      \multicolumn{4}{c}{} & \vbox{\vspace{4mm}} & 
      1      
    \end{array}
  \right)
  \,.
\end{equation}

\subsection{Combined Symmetry Breaking}
\label{sec:symbreakcombined}

Finally, we combine these results and choose a flat rank $2$ bundle
$W$ with
\begin{equation}
  \Hol(W) = 
  (\Z_3)_{(5)} \times (\Z_3)_{(3)}
  \,.
\end{equation}
It is clear from the $(\Z_3)_{(5)}$ and $(\Z_3)_{(3)}$ generators,
expressions~\eqref{eq:g5Z3} and~\eqref{eq:g3Z3basis} respectively,
that
\begin{equation}
  \label{eq:SMcommutant}
  H_{SU(4)\times (\Z_3)_{(5)}\times (\Z_3)_{(3)}} =
  SU(3) \times SU(2) \times U(1)_{(5)} \times U(1)_{(3)}
  \,.
\end{equation}
Proceeding sequentially, the $(\Z_3)_{(5)}$ part of $\Hol(W)$ leaves
$SU(5)\times U(1)_{(5)}$ invariant and decomposes the representation
of $\Spin(10)$ as in eq.~\eqref{eq:SU5reduce}. The second factor,
$(\Z_3)_{(3)}$, then breaks the $SU(5)$ to $SU(3)\times SU(2)\times
U(1)_{(3)}$ and decomposes the $\Rep{10}$, $\Rep{\overline{5}}$ and
$\Rep{1}$ of $SU(5)$ as
\begin{equation}
  \begin{split}
    \Rep{10} &= \Big( \Rep{3}, \Rep{2}, 1 \Big) \oplus \Big( \Rep{1},
    \Rep{1}, 3 \Big) \oplus \Big( \Rep{\overline{3}}, \Rep{1}, -1
    \Big)
    \\
    \Rep{\overline{5}} &= \Big( \Rep{1}, \Rep{2}, -3 \Big) \oplus
    \Big( \Rep{\overline{3}}, \Rep{1}, -1 \Big)
    \\
    \Rep{1} &= \Big( \Rep{1}, \Rep{1}, 3 \Big) \,.
  \end{split}
\end{equation}
The $SU(3)\times SU(2)$ representation content is precisely that of
the standard model $SU(3)_C \times SU(2)_L$. The physical hypercharge
and $B-L$ generators are then
\begin{equation}
  \begin{split}
    [Y] &= \frac{1}{2} [Y_{(5)}] + \frac{5}{6} [Y_{(3)}]
    \,,
    \\
    [Y_{B-L}] &= \frac{1}{3} [Y_{(3)}]
  \end{split}
\end{equation}
respectively. It follows that the low energy gauge group is
\begin{equation}
  H_{SU(4)\times (\Z_3)_{(Y)} \times (\Z_3)_{(B-L)}} =
  SU(3)_C \times SU(2)_L \times U(1)_Y \times U(1)_{B-L}
  \,,
\end{equation}
as desired.

Equation~\eqref{eq:pi1HolF} implies that to implement this symmetry
breaking pattern one must compactify on \CY{} threefolds $X$ with the
property that $(\Z_3)_{(Y)} \times (\Z_3)_{(B-L)} \subseteq \pi_1(X)$.
We will consider the minimal case where
\begin{equation}
  \label{eq:pi1XZ3Z3}
  \pi_1(X) = (\Z_3)_{(Y)} \times (\Z_3)_{(B-L)}
  \,.
\end{equation}
There are examples of such \CY{} threefolds as quotients of toric
hypersurfaces. However, in that framework we lack the necessary
understanding of stable holomorphic vector bundles. For this reason,
we are considering torus fibered \CY{} threefolds with fundamental
group eq.~\eqref{eq:pi1XZ3Z3}. These have not been presented
previously. In the remainder of this paper, we will give an explicit
construction of such a manifold and elucidate its properties.

\section{Torus Fibered \CY{} Threefolds}
\label{sec:CYintro}

A \CY{} threefold is a compact K\"ahler manifold $(X,J,g)$ of complex
dimension $3$ such that $\Hol(g)= SU(3)$. $J$ is the complex structure
and $g$ is the metric, which is then Ricci flat. That is, $X$ is a
manifold with a specific complex structure and K\"ahler metric, and
this metric has special holonomy. Such a threefold can be used to
compactify heterotic string theory to a $d=4$, $\Ncal=1$
supersymmetric vacuum.

In addition to requiring that $X$ be a \CY{} threefold, we will
demand two additional properties:
\begin{enumerate}
\item $X$ is torus fibered\footnote{This is sometimes called ``genus
    $1$ fibered''.}. That is, there exists an analytic map $\pi:X\to
  B$, where $B$ is a possibly singular complex surface and the
  generic fiber is a $T^2$.
  
  This is weaker than requiring an elliptic fibration, as a torus
  fibration does not necessarily have a section $s:B\to X$. In fact,
  our \CY{} threefold will not admit a section. However, a torus
  fibration is sufficient to construct stable holomorphic bundles.
\item $X$ has fundamental group $\pi_1(X)=\Z_3\times \Z_3$.
  
  As discussed in the previous section, such a fundamental group is
  large enough to break a $\Spin(10)$ gauge group via Wilson lines to
  the standard model group plus $U(1)_{B-L}$.
\end{enumerate}
There always exists the universal cover $\Xt$ of $X$, where, by
definition, $\pi_1(\Xt)=1$. Pulling back the metric $g$, we see that
$\Xt$ is a \CYm{} if $X$ is. The universal cover $\Xt$ comes with a
free\footnote{A free group action is one without fixed points.} group
action of $G\simeq \pi_1(X)$, the deck translations\footnote{The deck
  translations are sometimes also called covering automorphisms.}.
Conversely, given a simply connected \CY{} threefold $\Xt$ with a free
group action $G$ preserving the metric, then $G$ is a finite group and
it can be shown that the quotient $X\eqdef \Xt/G$ is a \CY{} threefold
with $\pi_1(X)=G$. So, a non-simply connected \CY{} threefold is
equivalent to a simply connected \CY{} threefold with a discrete
symmetry.

Therefore, we will construct an elliptically fibered \CY{} threefold
$\Xt$ with automorphism group $G=\Z_3\times\Z_3$. The base of the
elliptic fibration is a complex surface, which we choose (see
also~\cite{Schoen,SMbundles}) to be a rational elliptic surface
$dP_9$. This is not the only possibility, but is a useful choice
since $dP_9$ is itself elliptically fibered over $\CP^1$. Hence, the
full \CY{} threefold $\Xt$ admits a fibration over $\CP^1$
such that the generic fiber is the product of two elliptic curves. In
fact, such a threefold is automatically the fiber product of two
$dP_9$ surfaces over their base $\CP^1$, see~\cite{dP9torusfib}.

The fiber product is the ``universal'' way to fill in the pull back
diagram
\begin{equation}
  \vcenter{
    \xymatrix{
      \framebox{\vbox{\hbox{\hspace{6mm}\vspace{6mm}}}}
      \ar[d] \ar[r] & B_1\ar[d]_{\beta_1} \\
      B_2 \ar[r]^{\beta_2} & \CP^1
    }
  }
\end{equation}
arising from the projection of the two $dP_9$ surfaces to a common
base $\CP^1$. Explicitly, the fiber product is the hypersurface
\begin{equation}
  \Xt \eqdef
  B_1 \times_{\CP^1} B_2 \eqdef
  \Big\{ 
    (p_1,p_2) \in B_1\times B_2
  \Big|~
    \beta_1(p_1) = \beta_2(p_2)
  \Big\}
  \,.
\end{equation}
The fiber product of two $dP_9$ surfaces is simply connected. To
proceed, we must find a free $\Z_3\times \Z_3$ action on $\Xt$. The
quotient is then the desired \CY{} threefold $X$.

It is clear that this group action must preserve the fibration
structure. Otherwise the quotient \CYm{} would not be torus fibered.
Fortunately, this is automatic. Note that the anticanonical class of a
$dP_9$ is the class of the generic fiber $F$, see~\cite{dP9torusfib}.
That is,
\begin{equation}
  K_{dP_9} = -F
  \,.
\end{equation}
Now, any automorphism must preserve the canonical class, and, hence,
the fiber class. That is, the image of a generic fiber is an
irreducible curve homologous to a fiber. Consequently, it is again a
fiber.

To summarize, we want to consider the following:
\begin{itemize}
\item A \CY{} threefold $\Xt=B_1 \times_{\CP^1} B_2$, where $B_1$
  and $B_2$ are two rational elliptic surfaces ($dP_9$).
\item $G=\Z_3 \times \Z_3$ acts on each of $B_1$ and $B_2$ such that
  \begin{enumerate}
  \item The induced action on the base $\CP^1$ of $B_1$ and $B_2$ is
    the same. It then follows that the $G$ action extends to $\Xt$.
  \item The $G$ action on $\Xt$ is free.
  \end{enumerate}
\item $G$ acts non-trivially on the base $\CP^1$. This is not
  necessary, but yields more interesting actions on the homology. The
  case of trivial $\Z_3\times\Z_3$ action on the base $\CP^1$ is
  investigated in~\cite{Stienstra}, Proposition 7.1.
\end{itemize}
In the following section, we classify such rational elliptic surfaces.
It turns out that there exists a one-parameter family of such
surfaces. We will explicitly construct this family in
Section~\ref{sec:explicit}.

\section{Rational Elliptic Surfaces with Automorphisms}
\label{sec:surfaces}

\subsection{$\Z_3\times\Z_3$ Actions on Rational Elliptic Surfaces}
\label{sec:surfacesGaction}

Let $B$ be a \textdef{rational} \textdef{elliptic} surface, that is a
$dP_9$. First, recall the definition of such a surface. A surface is
fibered (over $\CP^1$) if there exists a projection map $\beta: B\to
\CP^1$. A fibered surface is elliptic if the generic fiber is a
complex torus and there exists a section, called the
\textdef{$0$-section}. In the following, we will always denote this
section by $\sigma$. A rational surface is a blow up of $\CP^2$ at a
finite number of points. If the surface is also elliptic in addition
to being rational, then the number of blowups must be $9$. This is the
origin of the subscript in ``$dP_9$''. Since each blowup increases the
Euler characteristic by $1$, the rational elliptic surface must have
\begin{equation}
  \chi( dP_9 ) = \chi(\CP^2) + 9 = 12
  \,.
\end{equation}
We can also compute the Euler characteristic from the elliptic
fibration point of view. A smooth fibration would be a $T^2$ bundle
and, hence,
\begin{equation}
  \chi( T^2 \text{ bundle over } \CP^1) = \chi(T^2) \chi(\CP^1) 
  = 0\cdot 2 = 0
  \,.
\end{equation}
Therefore, the elliptic fibration must degenerate somewhere. The
possible singular fibers were classified by Kodaira, and will be
reviewed at the beginning of Subsection~\ref{sec:singularfib}.
However, even without knowing the singular fibers explicitly, we can
immediately conclude that
\begin{equation}
  \label{eq:chidP9sum}
  \sum_{\text{singular fibers } F_s}
  \chi(F_s)
  = \chi(dP_9) = 12
  \,.
\end{equation}
Using the fact that a rational surface is simply connected, the Euler
characteristic determines the single nontrivial (co)homology group
\begin{equation}
  \label{eq:dP9cohomology}
  H_2(B,\Z) \simeq \Z^{10} \simeq H^2(B,\Z)
  \,.
\end{equation}

Furthermore, let $G=\Z_3 \times \Z_3$ act on $B$ such that it maps
fibers to fibers. For now, we will describe what one can conclude from
this characterization. We demonstrate the existence of such surfaces
in Section~\ref{sec:explicit}. Pick generators $g_1, g_2$ for $G$.
Abusing notation, also denote the corresponding automorphisms as
$g_{1,2}:B\to B$. By our assumptions, one of them (say, $g_1$) acts
non-trivially on the base $\CP^1$.  Therefore, we can choose
projective coordinates $s$, $t$ on $\CP^1$ such that the induced
action $\hat{g}_1$ on the base $\CP^1$ is
\begin{equation}
  \label{eq:g1}
  \hat{g}_1 = \beta \circ g_1 \circ \beta^{-1}
  :~ \CP^1\to\CP^1, \quad
  [s:t]\mapsto [s:\omega t]
  \qquad \qquad
  \omega \eqdef e^{\frac{2 \pi i}{3}}
  \,.
\end{equation}
Now, $\Aut(\CP^1)=PGL(2,\C)=PSL(2,\C)=SO^+(3,1)$ does not contain a
$\Z_3\times \Z_3$ subgroup. The reason is that a finite subgroup of
the proper, orthochronous Lorentz group $SO^+(3,1)$ cannot involve a
boost and, so, must lie in $SO(3)\subset SO^+(3,1)$. But $SO(3)$ does
not have a $\Z_3\times\Z_3$ subgroup as there is no platonic solid
with $9$ faces.

It follows that the $\Z_3\times\Z_3$ cannot act faithfully on the base
$\CP^1$. Thus, some combination of $\hat{g}_1$ and
$\hat{g}_2=\beta\circ g_2\circ \beta^{-1}$ must act trivially. We
will, without loss of generality, assume that
\begin{equation}
  \label{eq:g2}
  \hat{g}_2 = \Id_{\CP^1}:~ \CP^1\to\CP^1
  \,.
\end{equation}
Hence, $g_2$ leaves each fiber stable. In fact, we can fix the $g_2$
action uniquely. Pick a generic fiber $F$. Then $g_2|_F: F\to F$ is
some order $3$ automorphism of the elliptic curve $F$. There are $3$
possibilities for $g_2|_F$:
\begin{enumerate}
\item $g_2|_F = \Id_F$, but then $g_2$ would act trivially on $B$.
\item $F$ is the hexagonal torus\footnote{Tile the plane by regular
    hexagons. Then the vertices and midpoints form the hexagonal
    lattice. The quotient of the complex plane by the hexagonal
    lattice is the hexagonal torus. It is the only elliptic curve with
    a $\Z_3$ subgroup in the automorphism group.} and $g_2|_F$ is
  multiplication by $e^\frac{2\pi i}{3}$. In that case, the complex
  structure of the fiber $\beta^{-1}(\ptset)\simeq T^2$ in the
  elliptic fibration $\beta:B\to \CP^1$ must be constant. However, it
  can be shown that this is impossible.
\item $g_2|_F$ is the translation by a point of order $3$, that is, by
  a point $p\in F\simeq T^2$ satisfying\footnote{Here, by $\boxplus$
    we denote the addition of points on a torus.} $p\boxplus p\boxplus
  p=0$ (but $p\not=0$ and $p\boxplus p\not=0$). There are $8$ such
  points, see Figure~\ref{fig:threetorsion}.
  \begin{figure}[htbp]
    \centering \input{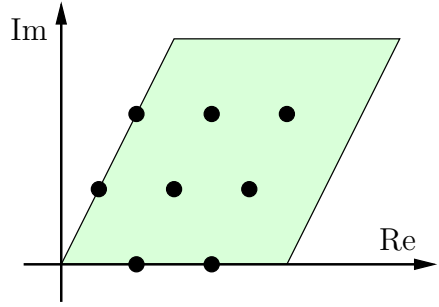}
    \caption{The $8$ points on $T^2$ of order $3$.}
    \label{fig:threetorsion}
  \end{figure}
\end{enumerate}
Everything except translation is ruled out, so $g_2$ must be the
translation by an order $3$ section of $B$, which we will call $\eta$.
That is,
\begin{equation}
  g_2 = t_\eta:~ B\to B
  \,.
\end{equation}
Now that we have determined the form of the $g_2$ action, let us
proceed to investigate $g_1$. There is another section defined by the
$G$ action on $B$, that is, the image of the $0$-section $\sigma$
under $g_1$. Denote this section by
\begin{equation}
  \xi \eqdef g_1(\sigma)
  \,.
\end{equation}
We can think of the $g_1$ action as first applying $t_{-\xi}\circ g_1$
(which fixes the $0$-section) and then translating by $\xi$,
\begin{equation}
  g_1 = t_{\xi} \circ \Big( t_{-\xi} \circ g_1 \Big)
  \,.
\end{equation}
In other words, we can factor the automorphism $g_1$ into $t_\xi$, the
translation by $\xi$, and another automorphism $\alpha_B$ which leaves
the $0$-section invariant. That is
\begin{equation}
  g_1 = t_\xi \circ \alpha_B:~ B\to B
\end{equation}
where
\begin{equation}
  \alpha_B \eqdef t_{-\xi} \circ g_1   
\end{equation}
and $\alpha_B(\sigma)=\sigma$.

\subsection{The \MWgrp}
\label{sec:MW}

The requirement that there be additional sections, besides $\sigma$,
restricts $B$ considerably. Let us review a few facts about sections.
First of all, a generic fiber of $B$ is a smooth elliptic curve.
Because we insist that $B$ has a section $\sigma$, there is a marked
point on the elliptic curve where $\sigma$ intersects the fiber. This
is precisely what is needed to define a group law on the curve. One
can think of an elliptic curve as a quotient $\C/\Lambda$ of the
complex plane and a lattice. The $0$-section fixes the origin of the
complex plane. The group law is simply the addition of complex
numbers, modulo the lattice $\Lambda$.

Hence, one can add sections to get another section just by adding the
points on each fiber. We will denote this addition of sections by
``$\boxplus$'', the same symbol as for the addition of points on a
single torus, to distinguish it from the addition of classes in
$H_2(B,\Z)$ which is denoted by ``$+$''. All sections form a countable
set, see~\cite{ShiodaMW}, so they compose a discrete Abelian group
under ``$\boxplus$'' called the \textdef{\MWgrp{}} $E(K)$.

Any section is, of course, a $2$-dimensional submanifold and, hence,
defines a class in $H_2(B,\Z)$. In fact, the section is uniquely
determined by its homology class. So we can think of $E(K)$ as a
subset of $H_2(B,\Z)$. However, it turns out that $E(K)$ can not, in
general, be a subgroup\footnote{To see that it is not a subgroup, note
  that for some rational elliptic surfaces $E(K)$ contains torsion
  elements while $H_2(B,\Z)$ is always torsion free.}. So our only
hope to describe the \MWgrp{} in terms of the homology group is to
write $E(K)$ as a quotient of $H_2(B,\Z)$. In fact, this is possible.
There is a subgroup $T\subset H_2(B,\Z)$ such that
\begin{equation}
  E(K) ~\simeq~ H_2(B,\Z) \Big/ T
  \,,
\end{equation}
as is proven in~\cite{ShiodaMW}. Moreover, the subgroup $T$ has a
rather simple description. $T$ is the subgroup generated by the
homology classes of
\begin{enumerate}
\item the $0$-section $\sigma$,
\item a generic fiber $F$, and
\item the irreducible components of singular fibers not intersecting
  $\sigma$. Note that the nonsingular fibers have only one irreducible
  component which must therefore intersect $\sigma$.
\end{enumerate}
The most natural description of the \MWgrp{} is, then, in terms of the
short exact sequence
\begin{equation}
  \label{eq:MWses}
  0 
  \longrightarrow T 
  \longrightarrow H_2(B,\Z) 
  \longrightarrow E(K) 
  \longrightarrow 0
  \,,
\end{equation}
which encodes $T$ as a subgroup and $E(K)$ as the corresponding
quotient of $H_2(B,\Z)$.

\subsection{The \MWlat}
\label{sec:MWlat}

There is more structure in the \MWgrp{} than just the addition law.
The intersection pairing in $H_2(B,\Z)$ induces the so-called
\textdef{height pairing}
\begin{equation}
  \label{eq:heightpairing}
  \left\langle \,\cdot\,, \,\cdot\, \right\rangle 
  :~
  E(K)\times E(K) \to \Q
\end{equation}
in the following way. Project the homology classes of the sections to
the orthogonal complement $T^\perp \subset H_2(B,\Q)$, and take their
intersection there. It is, in general, a rational number since the
projection does not always end up in $H_2(B,\Z)$. The \MWgrp{}
together with the (nondegenerate) height pairing forms the \MWlat.

We have the following explicit formula (see~\cite{ShiodaMW}) for the
height pairing of any two sections $\mu$, $\nu$. It is
\begin{equation}
  \label{eq:heightpairingformula}
  \langle \mu, \nu \rangle = 
  1 + \mu \sigma + \nu \sigma - \mu \nu 
  - \sum_{s\in \CP^1} \contr_s(\mu, \nu)
  \,,
\end{equation}
where multiplication of sections denotes the intersection product of
their homology classes and $\contr_s(\mu, \nu)$ is the entry of the
inverse of the intersection matrix $T_s$ associated with the
irreducible components not intersecting $\sigma$. Note that if the
fiber $\beta^{-1}(s)$ is smooth, then $T_s$ is a $0\times 0$ matrix.
Therefore, the sum in eq.~\eqref{eq:heightpairingformula} really only
runs over the points $s\in \CP^1$ where $\beta^{-1}(s)$ is a singular
fiber.

For example, assume that $\beta^{-1}(s)$ is an $I_3$ singular fiber of
the elliptic fibration. The $I_3$ singular fiber consists of three
\begin{figure}[htbp]
  \centering \input{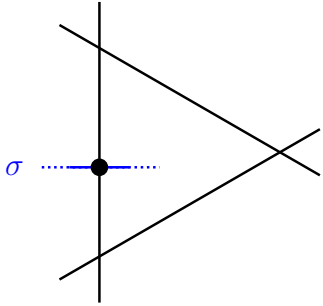}
  \caption{Singular fiber of the $I_3$ type.}
  \label{fig:I3fiberSimple}
\end{figure}
irreducible components, intersecting in a triangle. The $0$-section
$\sigma$ intersects one of the components, the so-called neutral
component. Thus, we are left with two components not intersecting the
$0$-section, see Figure~\ref{fig:I3fiberSimple}. Their intersection
matrix is the $A_2$ Cartan matrix with inverse
\begin{equation}
  \begin{pmatrix}
    2 & -1 \\ -1 & 2
  \end{pmatrix}^{-1} = 
  \begin{pmatrix}
    \frac{2}{3} & \frac{1}{3} \\ \frac{1}{3} & \frac{2}{3}
  \end{pmatrix}
  \,.
\end{equation}
Therefore, in this case, $\contr_s(\mu,\nu)$ can take the following
three values:
\begin{itemize}
\item $\contr_s(\mu,\nu)=0$ if either $\mu$ or $\nu$ intersect the
  neutral component.
\item $\contr_s(\mu,\nu)=\frac{2}{3}$ if both $\mu$ and $\nu$
  intersect the same component of the $I_3$ fiber and not the neutral
  component.
\item $\contr_s(\mu,\nu)=\frac{1}{3}$ if $\mu$, $\nu$ and $\sigma$
  each intersect a different component of the $I_3$ fiber.
\end{itemize}
It turns out that requiring $3$-torsion\footnote{Throughout this paper
  we use ``torsion'' in the group theoretical sense, that is, group
  elements of finite order: $g^n=1$ for some $n\in \Z$.} in the
\MWgrp{} is highly restrictive. Very few combinations of singular
fibers are possible, as will be discussed in the following section.

\subsection{Configurations of Singular Fibers}
\label{sec:singularfib}

Let us recall Kodaira's classification of singular fibers. If
$\beta^{-1}(s)$ is reducible, then the irreducible components not
intersecting the $0$-section form a root lattice (a sublattice of
$H_2(B,\Z)$) of the A-D-E type\footnote{The A-D-E lattices are the
  root lattices of the A-D-E Lie groups.}, which we will denote by
$T_s$. If $\beta^{-1}(s)$ is irreducible, then we write $T_s=()$, the
empty lattice. All possible singular fibers are listed in
Table~\ref{tab:Kodaira}.
\begin{table}[htbp]
  \centering
  \begin{tabular}{c|ccccccccc}
    Type of $\beta^{-1}(s)$ & 
    $I_1$ & $I_m, m\geq 1$ & 
    $II$ & $III$ & $IV$ & 
    $I_m^\ast$ & 
    $II^\ast$ & $III^\ast$ & $IV^\ast$
    \\ 
    $T_s$ &  
    $(\,)$ & $A_{m-1}$ &
    $(\,)$ & $A_1$ & $A_2$ &
    $D_{m+4}$ &
    $E_8$ & $E_7$ & $E_6$
    \\
    Euler characteristic & 
    $1$ & $m$ & $2$ & $3$ & $4$ & 
    $m+6$ & $10$ & $9$ & $8$
  \end{tabular}  
  \caption{Kodaira singular fibers.}
  \label{tab:Kodaira}
\end{table}
Note that a lattice $T_s$ either uniquely determines the Kodaira type
of the singular fiber, or is associated with a small number of
possible fibers. We can use this to find all possible configurations
of singular fibers that can occur when $E(K)$ has $3$-torsion.

For any point $s\in \CP^1$, we can decompose the fiber $\beta^{-1}(s)$
into irreducible components. These components determine homology
classes. Therefore, we have the embeddings
\begin{equation}
  \forall~s\in\CP^1:\quad
  T_s \subset H_2(B,\Z)
  \,.
\end{equation}
Of course, if $\beta^{-1}(s)$ is smooth, then the empty lattice
$T_s=()$ is embedded trivially. Since the fibers over different points
do not intersect, all $T_s$ must be embedded together as mutually
orthogonal sublattices
\begin{equation}
  \bigoplus_{s\in \CP^1} T_s ~\subset~ H_2(B,\Z)
  \,.
\end{equation}
The embeddings of sums of A-D-E root lattices $T_s$ in the $H_2(B,\Z)$
lattice are completely classified in~\cite{OguisoShioda}. Here, we
only quote their result. There are $74$ distinct sums
\begin{equation}
  \bigoplus_{s\in \CP^1} T_s \eqdef T ~\subset~ H_2(B,\Z)
  \,.
\end{equation}
For each of these, one can then determine the \MWgrp{} from
eq.~\eqref{eq:MWses}. This is a classical result which can also be
found in~\cite{OguisoShioda}. It turns out that only the $7$ lattice
embeddings given in Table~\ref{tab:MWthreetorsion} lead to
$3$-torsion.
\begin{table}[htbp]
  \centering
  \begin{tabular}{c||c|c|c|c|c|c|c}
    No. &
    39 & 
    51 & 
    61 & 
    63 & 
    66 & 
    68 & 
    69
    \\ \hline \vbox{\vspace{5mm}}
    $T$ &
    $A_2^{\oplus 3}$ & 
    $A_5\oplus A_2$ & 
    $A_2^{\oplus 3}\oplus A_1$ & 
    $A_8$ & 
    $A_5\oplus A_2 \oplus A_1$ & 
    $A_2^{\oplus 4}$ & 
    $E_6\oplus A_2$
    \\[1mm]
    $E(K)$ &
    $A_2^\ast \oplus \Z_3$ &
    $A_1^\ast \oplus \Z_3$ &
    $\langle\frac{1}{6}\rangle \oplus \Z_3$ &
    $\Z_3$ &
    $\Z_6$ &
    $\Z_3^2$ &
    $\Z_3$ 
  \end{tabular}  
  \caption{Possible \MWgrp{}s with $3$-torsion. In the first row we
    have also listed the case number
    of~\cite{OguisoShioda}. $\langle\frac{1}{6}\rangle$ denotes $\Z$ with
    the inner product matrix $(\frac{1}{6})\in \Mat(1,\Q)$.
  }
  \label{tab:MWthreetorsion}
\end{table}

It remains to list all possible configurations of singular fibers
corresponding to the $7$ possible lattices $T$. Each singular fiber
either is or is not one of the $g_1$-stable fibers. In the case where
the fiber is $g_1$-stable, it sits either over $[1:0]$ or $[0:1] \in
\CP^1$. Of course, nothing changes if we switch these two points, so
we can require that
$\chi\big(\beta^{-1}([1:0])\big)\geq\chi\big(\beta^{-1}([0:1])\big)$
without loss of generality. In the second case where the fiber is not
$g_1$-stable, the same singular fiber occurs $3$ times, cyclically
permuted by $g_1$.

We have found all consistent lattice embeddings. However, this is not
enough to infer the singular fibers. For example, we cannot determine
how many $I_1$ fibers there are, since this singular fiber does not
contribute to the $T$ lattice. The additional criterion we will use is
that the Euler characteristic of the singular fibers must add up to
$12$, see eq.~\eqref{eq:chidP9sum}. This fixes the singular fibers
uniquely except in case number 39 and 61 for which there are three
(39a,b,c) and two (61a,b) combinations respectively.  We
list\footnote{The attentive reader will notice that case 66 is
  missing. This is ruled out since $\oplus T_s = A_5\oplus A_2 \oplus
  A_1$ cannot come from a configuration of singular fibers. The three
  corresponding singular fibers cannot all sit over the two special
  points $[1:0]$, $[0:1] \in \CP^1$.} these combinations in
Table~\ref{tab:singularfib}.
\begin{table}[htbp]
  \renewcommand{\arraystretch}{1.3}
  \centering
  \begin{tabular}{c|ccccc}
    No. & $\oplus T_s$ & $E(K)$ & 
    $\beta^{-1}([1:0])$ & $\beta^{-1}([0:1])$ & 
    \parbox{2cm}{\centering Other sing. fibers}
    \\[3mm]
    \hline 
    39a & $A_2^{\oplus 3}$ & $A_2^\ast \oplus \Z_3$ &
    & & $3I_3$, $3I_1$
    \\ 
    (39b) & $A_2^{\oplus 3}$ & $A_2^\ast \oplus \Z_3$ &
    $II$ & $I_1$ & $3I_3$
    \\ 
    39c & $A_2^{\oplus 3}$ & $A_2^\ast \oplus \Z_3$ &
    & & $3IV$
    \\ 
    51 & $A_5\oplus A_2$ & $A_1^\ast \oplus \Z_3$ &
    $I_6$ & $I_3$ & $3I_1$
    \\
    61a & $A_2^{\oplus 3}\oplus A_1$ & 
    $\langle\frac{1}{6}\rangle \oplus \Z_3$ &
    $I_2$ & $I_1$ & $3I_3$
    \\
    (61b) & $A_2^{\oplus 3}\oplus A_1$ & 
    $\langle\frac{1}{6}\rangle \oplus \Z_3$ &
    $III$ & & $3I_3$
    \\
    63 & $A_8$ & $\Z_3$ &
    $I_9$ & & $3I_1$
    \\
    68 & $A_2^{\oplus 4}$ & $\Z_3^2$ &
    $I_3$ & & $3I_3$
    \\
    69 & $E_6\oplus A_2$ & $\Z_3$ &
    $IV^\ast$ & $IV$
  \end{tabular}
  \caption{Possible Configurations of Singular Fibers.}
  \label{tab:singularfib}
\end{table}

So far we only checked that the configuration of singular fibers gives
a consistent lattice embedding in $H_2(B,\Z)$ and has the right Euler
characteristic. However, not all combinations of singular fibers (with
Euler characteristic adding up to $12$) can occur in a rational
elliptic surface. Picking the singular fibers amounts to choosing the
monodromies of the smooth fibers as one goes around the singular
fiber. But not all choices of monodromy matrices are allowed since
encircling all singular fibers is a contractible loop. Hence, there
is a condition that the product of all monodromies is $1$.  In fact,
by comparing with the list of actually realized singular fibers
(see~\cite{Persson}), we can exclude cases 39b and 61b as well. This
is why they appear in brackets in Table~\ref{tab:singularfib}.

Finally, let us consider the dimension of the moduli space for these
rational elliptic surfaces. A generic $dP_9$ has $12$ singular fibers
and is uniquely determined by their position. Choosing coordinates on
the base $\CP^1$ amounts to fixing $3$ points. Hence, the moduli space
is $12-3=9$ dimensional. Of course, we are dealing with highly
non-generic $dP_9$ surfaces which admit $\Z_3\times\Z_3$
automorphisms. By choosing $g_1$ as in eq.~\eqref{eq:g1}, we have
already designated $2$ points to be the two fixed points. So the
position of one triple of singular fibers is the only remaining
freedom in the choice of the coordinate system. We see that the
surface in case 39a has one complex modulus, which is the position of
the second triple of singular fibers. All remaining surfaces are
isolated.  The number of parameters for each case is tabulated in
Table~\ref{tab:moduli}.
\begin{table}[htbp]
  \centering
  \begin{tabular}{c|ccccccc}
    No. &
    39a & 39c & 51 & 61a & 63 & 68 & 69
    \\ \hline
    Number of moduli & 
    $1$ & $0$ & $0$ & $0$ & $0$ & $0$ & $0$
  \end{tabular}
  \caption{Moduli for the surfaces with $3$-torsion in the \MWgrp.}
  \label{tab:moduli}
\end{table}

Furthermore, cases 39c, 63, and 68 are special points in the moduli
space of 39a. At these points, respectively, $I_1$ and $I_3$, $3I_3$,
and $3I_1$ fibers collide. This can be seen explicitly as a limit of
the Weierstrass model, see Appendix~\ref{sec:weierstrasslimit}.

\subsection{Actions with Restricted Fixed Point Sets}
\label{sec:fixedpoints}

Our goal is to construct free $G=\Z_3\times \Z_3$ group actions on the
fiber product $\Xt=B_1\times_{\CP^1} B_2$. We will show that not all
of the $dP_9$ surfaces in Table~\ref{tab:singularfib} can admit such a
group action. Rather, there is an additional restriction. Obviously,
the fiber over $s\in \CP^1$ contains a $\Z_3\subset G$ fixed point if
the fibers $\beta_{1}^{-1}(s)\subset B_1$ and
$\beta_{2}^{-1}(s)\subset B_2$ simultaneously have a $\Z_3$ fixed
point.  Conversely, if there are no fixed points for all choices of
$\Z_3\subset G$ then the action is free\footnote{$G=\Z_3\times \Z_3$
  is generated by $g_1$ and $g_2$. Therefore we have to check that
  $g_1$, $g_2$, and $g_1 g_2$ do not have fixed points.}.

First, let us review a standard argument showing that there must be at
least one fixed point on each rational elliptic surface $B$. This
shows that we will always have points $s\in \CP^1$ such that either
$\beta_1^{-1}(s)$ or $\beta_2^{-1}(s)$ contains a fixed point. The
argument is as follows.  For any fiber preserving automorphism
$1\not=K\in \Aut(B)$, the resolution\footnote{Blow up of the orbifold
  singularities.}  $\widehat{B/K}$ must again be a rational elliptic
surface.  This is a well known consequence of the classification of
surfaces. However, since it is an important part of our argument, we
review the proof it in Appendix~\ref{sec:resolutions}. Now assume $K$
acts without fixed points on $B$. Then, the $K$ quotient would be
smooth and
\begin{equation}
  12 = \chi\Big( \widehat{B/K} \Big) =
  \chi(B/K) = \frac{12}{|K|} < 12
  \,.
\end{equation}
This is a contradiction. Therefore there must be a fixed point of $K$
in $B$.

Consider the subgroup
\begin{equation}
  G_1\eqdef \{1,g_1,g_1^2\}  
  \,.
\end{equation}
By construction, this acts non-trivially on the base $\CP^1$. Of
course, the $G_1$ fixed points must then lie in the stable fibers
\begin{equation}
  F_0\eqdef \beta^{-1}([1:0])
  \,,\qquad 
  F_\infty\eqdef \beta^{-1}([0:1])
  \,.
\end{equation}
A simple application of the previous argument proves that there must
be a $G_1$ fixed point in either $F_0$ or $F_\infty$ in every case.
The same holds for the subgroup generated by $g_1g_2$, which also acts
non-trivially on the base $\CP^1$. Finally, consider the subgroup
\begin{equation}
  G_2\eqdef \{1,g_2,g_2^2\}  
  \,.
\end{equation}
Since $g_2$ is a non-zero translation, there are no $G_2$ fixed points
on smooth fibers. Counting the Euler characteristic of the resolution,
one can determine which singular fibers contain $G_2$ fixed points.
For example, in case 39a there is a $G_2$ fixed point in each of the
three $I_1$ fibers.

After these preliminaries, we can now discuss which $dP_9$ surfaces
can be used to construct fiber products $\Xt=B_1\times_{\CP^1} B_2$
with fixed point free $G_1$ actions. It turns out that the $G_2$
action can always be arranged to be fixed point free on the fiber
product, which is why we concentrate on $G_1$ in the following. We
naturally must distinguish between fixed points over $[1:0],[0:1] \in
\CP^1$ and the remaining fibers.

First, if $s\not= [1:0],[0:1] \in \CP^1$, then we can always rescale
(that is, act by the commutant of $\hat{g}_1$ in $PGL(2)$) the base
$\CP^1$ in, say, $B_2$ so that the singular fibers in $B_1$ and $B_2$
are not paired up in the fiber product. It follows that these fixed
points cannot prevent us from constructing a free action.

The only problematic case is for $s= [1:0]$ or $[0:1]$, since these
points are fixed by the $G_1$ action. We cannot rotate the relative
position of these points in $B_2$, because then the $G_1$ action would
no longer extend to the fiber product $\Xt$. Furthermore, we have
already seen that there must be a $G_1$ fixed point in at least one of
the two stable fibers $F_0$, $F_\infty$. This fiber must not be paired
in the fiber product with another fiber containing a $G_1$ fixed
point. Therefore, we can not use any surface that has a fixed point in
both $F_0$ and $F_\infty$. If the fibers are smooth tori, this poses no
additional restriction. As we have seen in
Figure~\ref{fig:threetorsion}, $\Z_3$ can act freely on $T^2$. Hence,
it is important to know which singular fibers allow free $\Z_3$
actions. It turns out to be quite difficult for $G_1$ to act freely on
the singular fibers, as we now discuss.

An obvious necessary condition for a Kodaira fiber to allow a free
$\Z_3$ action is that the Euler characteristic is divisible by $3$.
This leaves only $I_{3m}$, $III$, $I^\ast_{3m}$ and $III^\ast$.
Therefore, we can immediately rule out cases 61a and 69 since, in
these cases, neither of $F_0,F_\infty$ has suitable Euler
characteristic.

It is more difficult to exclude case 51. This will occupy the
remainder of this section. Any reader not interested in the details is
encouraged to jump to the summary at the end of this section. In case
51, the two fibers in question, $F_0=I_6$ and $F_\infty=I_3$,
certainly do have free $\Z_3$ automorphisms. Even so, the overall
topology of the rational elliptic surface $B$ prevents one from having
these special free automorphisms induced at $F_0$ and $F_\infty$. The
reason is again that for any fiber preserving automorphism $G\in
\Aut(B)$, the resolution $\widehat{B/G}$ must again be a rational
elliptic surface.
\begin{figure}[htbp]
  \centering \input{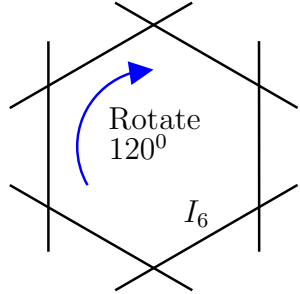}
  \caption{An automorphism of the $I_6$ singular fiber.}
  \label{fig:I6auto1}
\end{figure}
\begin{figure}[htbp]
  \centering \input{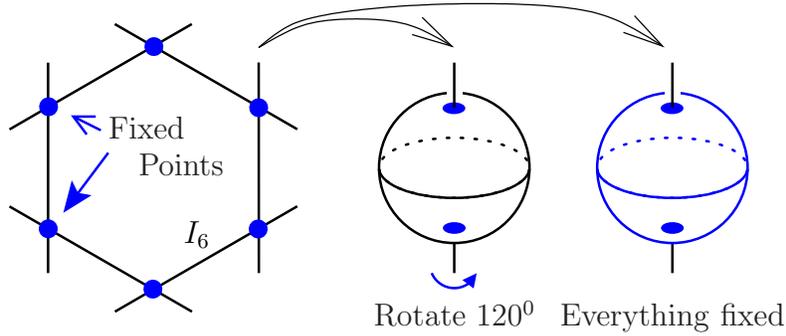}
  \caption{Other automorphisms of $I_6$. The $2$ possible actions on
    the irreducible components $(\CP^1)$ are indicated.}
  \label{fig:I6auto2}
\end{figure}
What do the possible $\Z_3$ automorphisms of $F_0$ and $F_\infty$ look
like? As an example, consider $F_0=I_6$. One possible automorphism is
the rotation of the hexagon, as depicted in Figure~\ref{fig:I6auto1}.
This is obviously a free $\Z_3$ action. Another possible automorphism
is shown in Figure~\ref{fig:I6auto2}. Here we act on each irreducible
component separately. In this case, there are obviously at least $6$
fixed points. The group of all possible automorphisms, $\Aut(I_6)$, is
then a semidirect product of these two types.

Now, assume that we are in case 51 and the only $G_1$ fixed points are
in the fiber $F_0$. There must be a fixed point, so $g_1|_{F_0}$
cannot be the automorphism of the $I_6$ depicted in
Figure~\ref{fig:I6auto1}.  This means that $g_1|_{F_0}: F_0\to F_0$
must fix at least the 6 intersection points of the irreducible
components. Depending on the action on the components themselves (see
Figure~\ref{fig:I6auto2}), the intersection points may or may not
become orbifold singularities. However, in either case, the fiber in
the resolution $\widehat{B/G_1}$ must be an $I_n$, $n\geq 6$. We can
determine $n$ by a simple Euler characteristic computation.
\begin{equation}
  12 = \chi\Big( \widehat{B/G_1} \Big) =
  \chi( I_n ) + \frac{1}{3}\Big( 3 \chi( I_1 ) + \chi(I_3) \Big)
  \quad \Rightarrow \quad n=10
  \,.
\end{equation}
This is a contradiction, since $I_{10}$ cannot appear as a singular
fiber in a rational elliptic surface. The rank $9$ lattice generated
by all the irreducible components except one, added to the lattice
generated by the $0$-section and the fiber, has overall rank $11$.
Therefore, it is too large to be embedded in the homology lattice
$H_2(B,\Z)\simeq \Z^{10}$. It remains to check that the other
distribution of fixed points in case 51, with all fixed points in
$F_\infty$, also can not occur.  By the same argument as above, the
resolution $\widehat{B/G_1}$ has singular fibers $I_9$, $I_2$ and
$I_1$. This is again excluded on dimensional grounds, since there are
again $9$ irreducible components (in $I_9$ and $I_2$) not intersecting
the $0$-section.

\begin{figure}[htbp]
  \centering \input{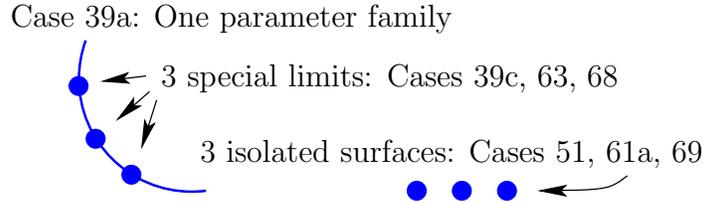}
  \caption{Hiker's guide to $dP_9$ with $\Z_3\times\Z_3$ action.}
  \label{fig:hiker}
\end{figure}

To summarize, we have found a classification of rational elliptic
surfaces with fiberwise $G=\Z_3\times\Z_3$ automorphisms which act
non-trivial on the base, see Figure~\ref{fig:hiker}. There is a single
$1$-parameter family of such surfaces. These surfaces are the only
ones that can be used to construct a free $G$ action on the fiber
product $\Xt$.  The three additional isolated surfaces will not lead
to fixed point free actions on the fiber product. Therefore, in the
remainder of the paper, we will focus on the generic case 39a.

\section{Explicit Realizations}
\label{sec:explicit}

\subsection{Weierstrass Generalities}
\label{sec:Wgeneral}

\subsubsection{The Weierstrass Equation}
\label{sec:Wequation}

Any elliptic surface $\beta:B\to \CP^1$ can be encoded in a
Weierstrass polynomial
\begin{equation}
\label{eq:Weierstrassgeneral}
  y^2 z = x^3 + a(t) x z^2 + b(t) z^3
  \,,
\end{equation}
where $x$, $y$, $z$ are homogeneous coordinates
\begin{equation}
  [x:y:z] = [\lambda x: \lambda y: \lambda z] 
  \quad \forall \lambda \not=0
\end{equation}
and $t$ is an affine coordinate on the base $\CP^1$.  For a rational
elliptic surface, $a(t)$ and $b(t)$ are polynomials in $t$ of degree
$4$ and $6$. Implicit in the Weierstrass equation, there are the
following conventions relating to the other coordinate chart $t \to
\overline{t} = \frac{1}{t}$.
\begin{gather}
  x \to \overline{x} \eqdef \frac{1}{t^2} x \,,\qquad y \to
  \overline{y} \eqdef \frac{1}{t^3} y \,,\qquad z \to \overline{z}
  \eqdef z
  \,,
  \\
  a \to \overline{a} \eqdef \frac{1}{t^4} a \,,\qquad b \to
  \overline{b} \eqdef \frac{1}{t^6} b \,.
\end{gather}
These rules are chosen so that after the coordinate change, one still
has a Weierstrass polynomial of the form
eq.~\eqref{eq:Weierstrassgeneral}.

These non-trivial transformation rules mean that $a$ and $b$ are not
functions but, rather, sections of the line bundles
\begin{equation}
  a \in \Gamma \Big(\Osheaf_{\CP^1}(4)\Big)
  \,,\qquad
  b \in \Gamma \Big(\Osheaf_{\CP^1}(6)\Big)
  \,,
\end{equation}
where $\Osheaf(n)$ is the sheaf of analytic functions homogeneous of
degree $n$. Similarly, $[x:y:z]$ is a section of the following bundle
$P$ of projective spaces over $\CP^1$:
\begin{equation}
  \xymatrix{
    \CP^2 \ar[r] & 
    P \eqdef \mathbb{P}\Big( 
    \Osheaf_{\CP^1}(2) \oplus 
    \Osheaf_{\CP^1}(3) \oplus 
    \Osheaf_{\CP^1} \Big)
    \ar[d]^p
    \\
    & 
    \CP^1
  }
  \,.
\end{equation}
The Weierstrass equation is then well defined on $P$, and we denote
its solution set by
\begin{equation}
  W_B \eqdef
  \Big\{ ([x:y:z],t) \in P 
  ~\Big|~
  \text{Weierstrass equation is satisfied}
  \Big \}
  \,.
\end{equation}
A fundamental fact is that, although they encode the same information,
in general
\begin{equation}
  B ~\not=~ W_B
  \,.  
\end{equation}
Note that a fiber of $W_B$ is a curve in the corresponding fiber of
$P$, that is, a cubic in $\CP^2$. But a cubic in $\CP^2$ has at most
$3$ irreducible components, whereas the singular fibers of $B$ may
contain up to $9$ irreducible components. Furthermore, $W_B$ is, in
general, a singular variety while $B$ is smooth.

This suggests that we have to resolve the singularities to identify
these spaces. Indeed,
\begin{equation}
  B = \widehat{W_B}
  \,.
\end{equation}
The advantage of the Weierstrass model is that it is very convenient
for computations. For example, it is easy to read off the singular
fibers of $B$ directly from the Weierstrass equation. The fibration
degenerates whenever the discriminant
\begin{equation}
  D(t)\eqdef 4 a(t)^3 + 27 b(t)^2
\end{equation}
vanishes. Since $D(t)$ is a polynomial of degree $12$ in $t$, it
follows that there are at most $12$ singular fibers.
\begin{table}[htbp]
  \centering
  \begin{tabular}{c|ccccccc}
    Kodaira fiber & 
    $\mathrm{order}\big(a(t-t_0)\big)$ & 
    $\mathrm{order}\big(b(t-t_0)\big)$ & 
    $\mathrm{order}\big(D(t-t_0)\big)$ 
    \\ \hline
    $I_0$ & any & any & $0$ \\
    $I_0$ & $\geq 3$ & $3$ & $6$ \\
    $I_0$ & $2$ & $\geq 4$ & $6$ \\
    $I_n$, $n\geq 1$ & $0$ & $0$ & $n$ \\
    $I_n$, $n\geq 1$ & $2$ & $3$ & $n+6$ \\
    $I_0^\ast$ & $2$ & $3$ & $6$ \\
    $II$       & $\geq 1$ & $1$ & $2$ \\
    $IV^\ast$  & $\geq 3$ & $4$ & $8$ \\
    $III$      & $1$ & $\geq 2$ & $3$ \\
    $III^\ast$ & $3$ & $\geq 5$ & $9$ \\
    $IV$       & $\geq 2$ & $2$ & $4$ \\
    $II^\ast$  & $\geq 4$ & $5$ & $10$ \\
  \end{tabular}
  \caption{Singular fibers of the Weierstrass equation at $t=t_0$.}
  \label{tab:weierstrassdegrees}
\end{table}
The precise nature of a singular fiber can then be read off from the
order of vanishing of $D$, $a$ and $b$ at that fiber. The requisite
information is presented in Table~\ref{tab:weierstrassdegrees},
see~\cite{Bershadsky:1996nh}.

\subsubsection{Sections}
\label{sec:Wsections}

Another advantage of the Weierstrass model is that one can describe
sections explicitly. An arbitrary section can be written as
\begin{equation}
  \rho: \CP^1 \to W_B, \quad 
  t \mapsto [\rho_x(t): \rho_y(t): \rho_z(t)]
  \,.
\end{equation}
Hence, it is defined by three polynomials $\rho_x$, $\rho_y$, and
$\rho_z$ satisfying the Weierstrass equation. Of course, globally,
$\rho_x$, $\rho_y$, and $\rho_z$ are sections of some sheaves.
Therefore, they transform like $x$, $y$, and $z$ as we change the
coordinate patch. As long as $\rho_z$ is not identically zero, one can
use the homogeneous rescaling and write the section as
\begin{equation}
  \rho: \CP^1 \to W_B, \quad 
  t \mapsto [\tilde{\rho}_x(t): \tilde{\rho}_y(t): 1]
  \,,
\end{equation}
where $\tilde{\rho}_x$ and $\tilde{\rho}_y$ are now locally meromorphic
functions. Globally, they must be sections of the sheaf of meromorphic
functions of homogeneous degree $2$ and $3$. That is,
\begin{equation}
  \tilde{\rho}_x\in \Gamma\Big(\Msheaf(2)\Big)
  , \qquad
  \tilde{\rho}_y\in \Gamma\Big(\Msheaf(3)\Big)
  \,.
\end{equation}
In all cases, the section
\begin{equation}
  \label{eq:WeierstrassZero}
  \sigma: \CP^1 \to B,\quad t\mapsto [0:1:0]
\end{equation}
always exist and is, by convention, declared to be the $0$-section.

Similarly, the addition law has a clear geometric meaning. Recall that
to define the addition law for sections, one need only specify the
addition of points in each fiber. By continuity, we only have to do
this for the generic (smooth) fibers. In the Weierstrass model, these
are simply smooth cubic curves in $\CP^2$. Explicitly, fix $t=t_0$ and
define the curve
\begin{equation}
  C \eqdef 
  \Big\{
    [x:y:z] 
  ~\Big|~
    y^2 z = x^3 + a(t_0) x z^2 + b(t_0) z^3    
  \Big\}
  \subset \CP^2
  \,.
\end{equation}
Together with the chosen origin
\begin{equation}
  0 \eqdef [0:1:0] \in C
  \,,
\end{equation}
this is an Abelian group.

To define the group law, we have to specify the sum $p_1\boxplus p_2$
for any two points $p_1,p_2\in C$. However, for our purposes, it is
more useful to give the group law in a more symmetric way. For any two
points $p_1$ and $p_2$, one must specify a unique $p_3$ such that
\begin{equation}
  p_1 \boxplus p_2 \boxplus p_3 = 0
  \,.
\end{equation}
The group law is then as follows. Any line
\begin{equation}
  L
  \eqdef
  \Big\{
    [x:y:z] 
  ~\Big|~
    \ell_1 x + \ell_2 y + \ell_3 z = 0
  \Big\}
  \subset \CP^2
  \,,\quad
  (\ell_1,\ell_2,\ell_3)\in \C^3-\{0\}
\end{equation}
intersects the cubic $C$ in three points
\begin{equation}
  \Big\{ p_1,\, p_2,\, p_3 \Big\} \eqdef C \cap L
  \,,
\end{equation}
possibly with multiplicity. By definition, these points add up to
zero,
\begin{equation}
  p_1 \boxplus p_2 \boxplus p_3 = 0
  \,.
\end{equation}
It is clear that $p_3$ is unique, since there is only a single line
$L$ passing through $p_1$ and $p_2$. One can geometrically check that
this addition law satisfies the group axioms.
\begin{figure}[htbp]
  \centering \input{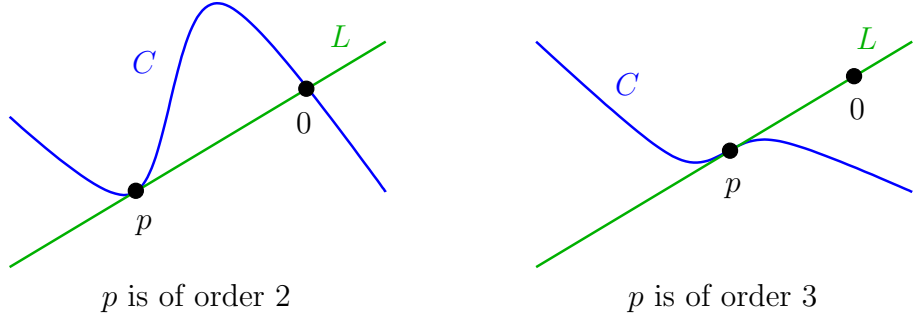}
  \caption{Special cases of the geometric group law.}
  \label{fig:tangentflex}
\end{figure}

Because we allow for multiplicities, that is, $L$ may be tangent to
$C$, there are the following two important special cases (see
Figure~\ref{fig:tangentflex}). First, $p\in C$ is of order $2$ if
$p\boxplus p\boxplus 0=0$. The line $L$ is then tangent to $C$ at $p$
and intersects $C$ transversely at $0$. Second, if $p$ is a point of
order $3$ then $p\boxplus p\boxplus p=0$. Geometrically, this means
that $L$ intersects $C$ in a flex. A smooth cubic always has $9$
flexes, one being $[0:1:0]$ in the Weierstrass form.

\subsection{Weierstrass Model}
\label{sec:weierstrass}

Although we know from the above classification that surfaces with
suitable singular fibers and a $3$-torsion section exist, it is not
obvious that they actually have a $G=\Z_3\times \Z_3$ action. One way
to ensure this is to write down an explicit realization. Consider the
following Weierstrass equation
\begin{equation}
  \label{eq:weierstrass}
  y^2 z = x^3 + 
  \left[ t+t^4\left( \gamma + \frac{1}{48} \right) \right] x z^2 +
  \left[ 1+2 \gamma t^3 + \left( \gamma^2-\frac{1}{1728} 
    \right)  t^6 \right] z^3
  \,,
\end{equation}
where $t$ is the coordinate on the base $\CP^1$ and $\gamma$ is the
one complex parameter of the family. This turns out to correspond to
case 39a in Table~\ref{tab:singularfib}. First, we check that it has
the correct configuration of singular fibers. The discriminant is
\begin{equation}
  \label{eq:disc}
  D = 27 \left[ 1+ \left(\gamma+\frac{1}{24} \right)t^3\right]^3 
  \left[ 1+ \left(\gamma+\frac{5}{216}\right) t^3 \right]
  \,.
\end{equation}
From this factorization and Table~\ref{tab:weierstrassdegrees}, we can
easily read off the singular fibers. For generic $\gamma$, there is an
$I_3$ at each of the three roots of $t^3=-(\gamma+\frac{1}{24})^{-1}$
and an $I_1$ at each of the three roots of
$t^3=-(\gamma+\frac{5}{216})^{-1}$.

The generators of $G=\Z_3\times \Z_3$ are $g_1=t_\xi \circ \alpha_B$
and $g_2=t_\eta$, so we have to describe $\xi$, $\eta$ and $\alpha_B$.
First, $\alpha_B$ is the order $3$ automorphism
\begin{equation}
  \label{eq:alpha}
  \alpha_B:B\to B, \quad
  \big([x:y:z],\,t\big) \mapsto 
  \big([\omega x: y:z], \,\omega^{-1}t\big)
  \qquad 
  \omega \eqdef e^{\frac{2\pi i}{3}}
  \,.
\end{equation}
It obviously preserves the Weierstrass equation
eq.~\eqref{eq:weierstrass} as well as the $0$-section,
eq.~\eqref{eq:WeierstrassZero}. The order $3$ section is
\begin{equation}
  \label{eq:eta}
  \eta: \CP^1 \to B, \quad 
  t \mapsto \left[ \frac{1}{12} t^2 : 
    1+\left(\gamma+\frac{1}{24}\right) t^3  : 1 \right]
  \,.
\end{equation}
It is straightforward to check that $\eta$ is of order $3$, that is
\begin{equation}
  \eta\boxplus \eta \boxplus \eta = \sigma
  \,,
\end{equation}
or, equivalently, $\eta(t)$ is a flex for all $t$. Furthermore, $\eta$
is preserved by $\alpha_B$. Hence, $g_1$ and $g_2$ commute.

Finally, we should give the section $\xi$ satisfying
\begin{equation}
  \label{eq:xiMWeq}
  \Big( t_\xi \circ \alpha_B \Big)^3 = 1
  \quad \Leftrightarrow \quad
  \alpha_B^2 \xi \boxplus \alpha_B \xi \boxplus \xi = 0
  \,.
\end{equation}
Any general equation for the section $\xi$ is exceedingly complicated
since the section must have monodromies around the special values for
$\gamma$. Therefore, the coefficients of the polynomials $\xi_x(t)$,
$\xi_y(t)$ must contain roots of polynomials in $\gamma$.  To simplify
matters, we will choose a specific value $\gamma=-\frac{1}{48}$. Note
that this value is not one of the special points corresponding to
cases 39c, 63, 68. However, it simplifies the Weierstrass equation
somewhat, which now becomes
\begin{equation}
  \label{eq:specialweierstrass}
  y^2 z = x^3 + t x z^2 +
  \left[ 1-\frac{1}{24} t^3 -\frac{1}{6912} t^6 \right] z^3
  \,.
\end{equation}
For this special value we can now write down the section $\xi$. Pick
one root of $r^3 = \frac{\sqrt{3}}{96}i$. Then, the following is a
section
\begin{equation}
  \label{eq:xi}
  \xi: \CP^1 \to B, \quad 
  t \mapsto \left[ 
    \frac{\sqrt{3}}{6}i \left(4 r t + \frac{1}{r} \right)  : 
    \frac{1}{\sqrt{3}}i \left(1 + \frac{1}{48} t^3 \right) : 
    1 \right]
  \,.
\end{equation}
One immediately recognizes that the $\alpha_B$-action on $\xi$
corresponds precisely to the choice of the $3$rd root for $r$.
Moreover, since only one coordinate depends on $r$, the three
$\alpha_B$-images lie on a line. That is, the points add to $0$ on
each fiber. Hence eq.~\eqref{eq:xiMWeq} is satisfied. For other
nondegenerate values of $\gamma$, one can also find sections since the
\MWgrp{} must remain the same. Furthermore, since the
$\alpha_B$-action on the discrete \MWgrp{} has to be invariant under
smooth deformations, the section likewise solves eq.~\eqref{eq:xiMWeq}.

\subsection{Pencil of Cubics}
\label{sec:pencil}

Another way to describe a rational elliptic surface is as a pencil of
cubics. For completeness, we describe our $dP_9$ in this framework.

A pencil of cubics is given by two homogeneous cubic polynomials
$F(x,y,z)$ and $G(x,y,z)$.  Their ratio $[F:G]\in \CP^1$ is well
defined apart from the $9$ points\footnote{For simplicity, we assume
  here that the $9$ points are distinct. This is the case in the
  pencil we are interested in.}  where $F=G=0$. Hence, their ratio is
not quite a function, but rather a rational\footnote{Rational maps are
  customarily denoted by a broken arrow $A\dashrightarrow B$.} map
$[F:G]:~ \CP^2 \dashrightarrow \CP^1$. However, after blowing up these
$9$ points in $\CP^2$, we do get a map that is defined everywhere,
\begin{equation}
  \mathrm{Bl}(\CP^2) \longrightarrow \CP^1
  \,.
\end{equation}
This blowup is our rational elliptic surface. Put differently,
consider the hypersurface
\begin{equation}
  B \eqdef 
  \left\{ \Big( [x:y:z],~ [\mu:\nu] \Big) 
  \Big|~ \mu F(x,y,z) + \nu G(x,y,z) = 0 \right\}
  \subset \CP^2 \times \CP^1
  \,.
\end{equation}
Projecting to the $\CP^2$ factor of the ambient space, we see that $B$
is a blow-up of $\CP^2$ at $9$ points. Projecting to the $\CP^1$, we
see that $B$ is elliptically fibered.

For example, take
\begin{equation}
  \label{eq:specialcubicpencil}
  \begin{split}
    F(x,y,z) =&~ \sqrt[3]{2} \Big( x^2 y+\omega y^2 z+\omega^2 z^2 x
    \Big) , \qquad \omega \eqdef e^\frac{2 \pi i}{3}
    \\
    G(x,y,z) =&~ -\frac{3+i\sqrt{3}}{36} \Big(x^3+y^3+z^3+6 x y z\Big)
    \,.
  \end{split}
\end{equation}
The Weierstrass form of the cubic $F(x,y,z)+tG(x,y,z)$ is
precisely eq.~\eqref{eq:specialweierstrass}, corresponding to
$\gamma=-\frac{1}{48}$ in our one parameter family of rational
elliptic surfaces. This pencil of cubics has $9$ distinct base points,
which is to say that there are $9$ solutions to $F=G=0$, all with
multiplicity $1$.

One of the $\Z_3\times \Z_3$ generators is straightforward to write
down:
\begin{equation}
  \label{eq:g1pencil}
  g_1: \CP^2\times \CP^1 \to \CP^2\times\CP^1,
  \Big( [x:y:z],~ [\mu,\nu] \Big) \mapsto
  \Big( [y:z:x],~ [\mu,\omega \nu] \Big)
  \,.
\end{equation}
It preserves the hypersurface, $g_1(B)=B$, and, therefore, restricts
to a $\Z_3$ action on $B$. Of course, we know that there must be
another $\Z_3$ automorphism of $B$ which commutes with
eq.~\eqref{eq:g1pencil}.  However, within the description by a pencil
of cubics, it is difficult to discuss the sections of the elliptic
fibration. Recall that the $\Z_3$ generator is a translation by an
order $3$ section.  Therefore, we must be able to explicitly write
down the sections and the group action on them.  This is much easier
to do in the Weierstrass model.

\section{Homology of the Surface}
\label{sec:dP9homology}

\subsection{Geometry of Sections and Singular Fibers}
\label{sec:MWweierstrass}

Now that we have established the existence of a rational elliptic
surface $B$ with $G=\Z_3\times \Z_3 \subset \Aut(B)$, we want to
compute the action of $G$ on the homology group $H_2(B,\Z)$. Using
this, we can compute the Hodge numbers of our \CYm{}. However, first
we have to find one final piece of information.  That is, we need to
know how the sections $\sigma$, $\xi$, $\eta$, and the singular fibers
intersect.

To do this, consider a section $\rho: \CP^1\to B,~ t\mapsto
[\rho_x(t):\rho_y(t):1]$. Again, globally, $\rho_x\in
\Gamma\big(\Msheaf(2)\big)$ and $\rho_y\in
\Gamma\big(\Msheaf(3)\big)$, that is, they are meromorphic functions
of homogeneous degree $2$ and $3$ respectively. In general, $\rho_x$
and $\rho_y$ will have poles. There is nothing wrong with this since
they correctly define homogeneous coordinates in $\CP^2$. A pole only
signifies that one should do a homogeneous rescaling, in which case the
homogeneous coordinates will be finite. For certain sections, $\rho_x$
and $\rho_y$ will have no poles. In this case, $\rho_x$ and $\rho_y$
are actual polynomials of degree $2$ and $3$ respectively and
homogeneous rescalings are not necessary. This implies that
\begin{equation}
  [\rho_x(t):\rho_y(t):1] \not= [0:1:0] 
  \qquad \forall t \in \CP^1
  \,.
\end{equation}
In other words, such sections do not intersect the $0$-section
$\sigma$. In fact, the converse is true. These are precisely the
sections not intersecting $\sigma$. There are at most $240$ such
sections, their height pairing is $\langle \rho,\rho \rangle \leq 2$
and they generate the \MWgrp. This follows again from the
classification of A-D-E lattices and is proven in~\cite{ShiodaMW}.

\begin{figure}[htbp]
  \centering \input{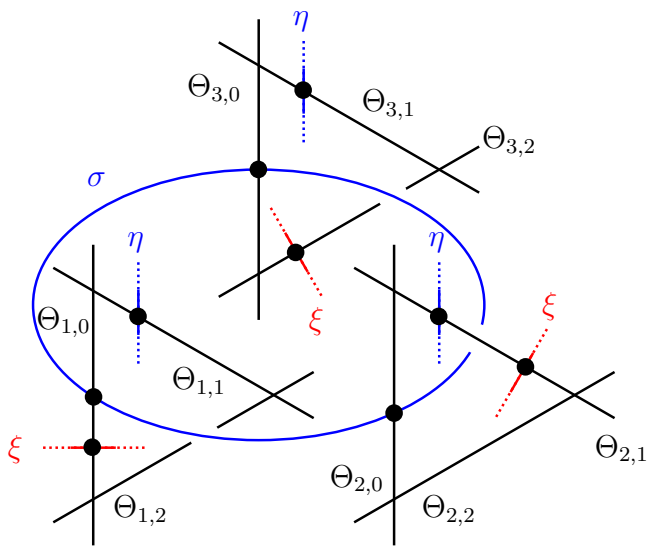}
  \caption{Sections and reducible fibers on $B$.}
  \label{fig:sectionreducible}
\end{figure}

From the eqns.~\eqref{eq:xi} and~\eqref{eq:eta}, it is obvious that
the sections $\eta$, $\xi$ do not intersect $\sigma$. It remains to
understand where they intersect the singular fibers. Once this is
known, we can compute their height pairings,
eq.~\eqref{eq:heightpairingformula}. In our case\footnote{That is, the
  rational elliptic surface defined by the Weierstrass equation
  eq.~\eqref{eq:weierstrass}. Here, of course, there are fewer than
  the aforementioned $240$ sections not intersecting $\sigma$.}, the
only singular fibers with multiple irreducible components are the three
$I_3$ fibers.
\begin{figure}[htbp]
  \centering \input{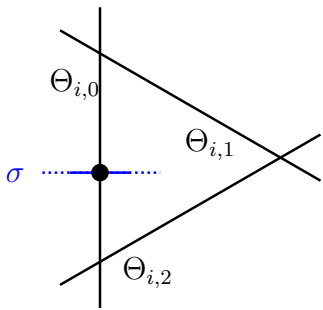}
  \caption{The $i^\mathrm{th}$ singular fiber of $I_3$ type.}
  \label{fig:I3fiber}
\end{figure}
Let $\Theta_{i,0}$ be the reducible component of the $i^{\text{th}}$
$I_3$ singular fiber which intersects $\sigma$, where $i=1,2,3$.
Denote by $\Theta_{i,1}$ and $\Theta_{i,2}$ the other reducible
components. The structure of the $i^\text{th}$ singular $I_3$ fiber is
shown in Figure~\ref{fig:I3fiber}. Using eq.~\eqref{eq:xi} for the
section $\xi$, we can easily check that it passes through the
singularity for $2$ of the $3$ $I_3$ singular fibers. (The surface $B$
is smooth, but the Weierstrass model $W_B$ contains singularities).
Likewise, we can use eq.~\eqref{eq:eta} to show that $\eta$ passes
through all $3$ singular points, and so, never intersects
$\Theta_{i,0}$, $i=1,2,3$.  Hence, the curves must intersect as
in Figure~\ref{fig:sectionreducible}, up to relabeling of
$\Theta_{i,j}$.

\begin{figure}[htbp]
  \centering \input{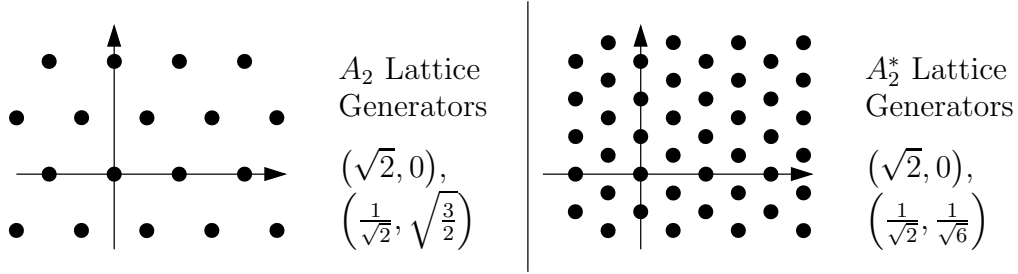}
  \caption{The $A_2$ and $A_2^\ast$ lattice.}
  \label{fig:A2lattice}
\end{figure}

We can summarize the structure of the \MWlat{} as follows. The free
part of the \MWlat{} is $A_2^\ast$, the dual of the $A_2$ root lattice
(see Figure~\ref{fig:A2lattice}).
This lattice is generated by $2$ points of minimal length
$(\frac{2}{3})^{\frac{1}{2}}$. Actually, the $6$ minimal lattice
points are nothing else but the section $\xi$ of eq.~\eqref{eq:xi} and
its images under $\alpha_B$ and $(-1)_B$ (the involution on $B$ acting
as $-1$ in each fiber separately). All height pairings can be simply
computed from the lattice, Figure~\ref{fig:A2lattice}.

To check this, first note that these $6$ sections are indeed distinct
and of the same height norm since the height pairing is preserved by
$\alpha_B$ and $(-1)_B$. We have only to calculate $\langle
\xi,\xi\rangle$. This can be done using the general formula in
eq.~\eqref{eq:heightpairingformula}. From this, we can immediately
compute $\langle \xi,\xi\rangle$ to be
\begin{equation}
  \label{eq:heightpairingxi}
  \langle \xi, \xi \rangle = 
  1 + 0 + 0 - (-1) 
  - \begin{pmatrix}
    2 & -1 \\ -1 & 2
  \end{pmatrix}^{-1}_{(1,1)}
  - \begin{pmatrix}
    2 & -1 \\ -1 & 2
  \end{pmatrix}^{-1}_{(2,2)}
  = \frac{2}{3}
  \,.
\end{equation}
Hence, $\xi$ really is a minimal length lattice point. Finally, note
that although we used the explicit equation for the sections to keep
things simple, the intersection pattern in
Figure~\ref{fig:sectionreducible} is required by the height paring of
$\xi$, $\eta$, and $\xi\boxplus \eta$ whenever $\xi$ is a minimal
length point of the \MWlat. This is explained in more detail in
Appendix~\ref{sec:synthetic}.

\subsection{Action on Homology}
\label{sec:surfaceHaction}

To determine the $G=\Z_3\times\Z_3$ action on the homology group
$H_2(B,\Z)$, we first have to pick convenient generators. To do this,
recall the description of the \MWgrp{} via the exact sequence
\begin{equation}
  0 
  \longrightarrow T 
  \longrightarrow H_2(B,\Z) 
  \longrightarrow E(K) 
  \longrightarrow 0
  \,.
\end{equation}
We can turn this description around and think of $H_2(B,\Z)$ as an
extension of $T$ by $E(K)$. This implies that we can generate all of
$H_2(B,\Z)$ using images of generators of $T$ and the lifts of
generators of $E(K)$. The generators of $T$ are naturally generators
of the homology group, and the obvious lift of the sections is to take
their homology classes\footnote{Warning: for the rest of this paper,
  we make no distinction between a section and its homology class.}.

Hence, we choose the following set of generators for the homology
group of our rational elliptic surface.
\begin{descriptionlist}
\item[$\sigma$:] The $0$-section
\item[$F$:] A generic fiber.
\item[$\Theta_{1,1},\dots,\Theta_{3,2}$:] The 6 irreducible components
  of the three $I_3$ singular fibers not intersecting $\sigma$. We
  choose the indexing such that $\alpha_B$ cyclically permutes
  $\Theta_{1,i}\to\Theta_{2,i}\to \Theta_{3,i}$.
\item[$\xi,\alpha_B \xi$:] Two sections generating the free part of
  the \MWgrp.
\end{descriptionlist}
These form a basis for $H_2(B,\Q)$, but they only generate an index
$3$ sublattice of $H_2(B,\Z)$ as is clear from the short exact
sequence above. To generate all of $H_2(B,\Z)$, we have to include the
homology class of the generator of $E(K)_\Tor$.
\begin{descriptionlist}
\item[$\eta$:] The $\alpha_B$-invariant order $3$ section $\eta$.
\end{descriptionlist}
Of course, we have now chosen $11$ generators. Since $\rank
H_2(B,\Z)=10$, there must be one relation between these. It is given
by
\begin{equation}
  \label{eq:homologyrelation}
  \eta = \sigma + F 
  - \frac{2}{3} \Big( \Theta_{1,1}+\Theta_{2,1}+\Theta_{3,1} \Big)
  - \frac{1}{3} \Big( \Theta_{1,2}+\Theta_{2,2}+\Theta_{3,2} \Big)
  \,.
\end{equation}
This relation can easily be checked by computing intersection
numbers\footnote{The only possible complication is finding the
  intersection of the left side with $\xi$. This follows from the
  height pairing $\langle \eta,\xi\rangle=0$ and
  eq.~\eqref{eq:heightpairingformula}.} using
Figure~\ref{fig:sectionreducible}. The intersection number of both
sides with the $10$ generators of $H_2(B,\Q)$ is the same. Since there
is no torsion in the integral homology group $H_2(B,\Z)$, this implies
the equality eq.~\eqref{eq:homologyrelation}. We could now eliminate
one generator, for example $\Theta_{1,2}$, and have a basis for
$H_2(B,\Z)$. But then we would break the symmetry between the three
$I_3$ fibers. Hence, we prefer to work with the $11$ generators, even
though one of them is redundant. For reference, we list their
intersection matrix in Table~\ref{tab:intersection}.
\begin{table}[htbp]
  \centering
  \begin{tabular}{c|@{\hspace{5mm}}c@{\hspace{5mm}}cccccccccc}
    $\cdot$ & 
    $\sigma$ & $F$ &
    $\Theta_{1,1}$ & $\Theta_{2,1}$ & $\Theta_{3,1}$ &
    $\Theta_{1,2}$ & $\Theta_{2,2}$ & $\Theta_{3,2}$ &
    $\xi$ & $\alpha_B \xi$ & $\eta$ 
    \\ \hline
    $\sigma$ & 
    $-1$ & $1$ & 
    $0$ & $0$ & $0$ & $0$ & $0$ & $0$ & 
    $0$ & $0$ & $0$ 
    \\
    $F$ &
    $1$ & $0$ & 
    $0$ & $0$ & $0$ & $0$ & $0$ & $0$ & 
    $1$ & $1$ & $1$ 
    \\
    $\Theta_{1,1}$ & 
    $0$ & $0$ & 
    $-2$ & $0$ & $0$ & $1$ & $0$ & $0$ & 
    $0$ & $0$ & $1$ 
    \\
    $\Theta_{2,1}$ & 
    $0$ & $0$ & 
    $0$ & $-2$ & $0$ & $0$ & $1$ & $0$ & 
    $1$ & $0$ & $1$ 
    \\
    $\Theta_{3,1}$ &
    $0$ & $0$ & 
    $0$ & $0$ & $-2$ & $0$ & $0$ & $1$ & 
    $0$ & $1$ & $1$ 
    \\
    $\Theta_{1,2}$ & 
    $0$ & $0$ & 
    $1$ & $0$ & $0$ & $-2$ & $0$ & $0$ & 
    $0$ & $1$ & $0$ 
    \\
    $\Theta_{2,2}$ & 
    $0$ & $0$ & 
    $0$ & $1$ & $0$ & $0$ & $-2$ & $0$ & 
    $0$ & $0$ & $0$ 
    \\
    $\Theta_{3,2}$ &
    $0$ & $0$ & 
    $0$ & $0$ & $1$ & $0$ & $0$ & $-2$ & 
    $1$ & $0$ & $0$ 
    \\
    $\xi$ & 
    $0$ & $1$ & 
    $0$ & $1$ & $0$ & $0$ & $0$ & $1$ & 
    $-1$ & $1$ & $0$ 
    \\
    $\alpha_B \xi$ & 
    $0$ & $1$ & 
    $0$ & $0$ & $1$ & $1$ & $0$ & $0$ & 
    $1$ & $-1$ & $0$ 
    \\
    $\eta$ &
    $0$ & $1$ & 
    $1$ & $1$ & $1$ & $0$ & $0$ & $0$ & 
    $0$ & $0$ & $-1$ 
  \end{tabular}
  \caption{Intersection matrix of the homology generators.}
  \label{tab:intersection}
\end{table}

Now that we have defined a set of generators for the homology group,
we want to know how $G$ acts on them. That is, we must determine the
push forwards $(\alpha_B)_\ast$, $(t_\xi)_\ast$, and $(t_\eta)_\ast$.
Most follow directly from the definition. The only tricky part is to
find the action of $(t_\xi)_\ast$ and $(t_\eta)_\ast$ on the sections
$\xi$, $\alpha_B\xi$, and $\eta$. In other words, we have to determine
the homology classes of $\xi\boxplus \xi$, $\xi\boxplus\alpha_B \xi$,
$\eta\boxplus\xi$, $\eta\boxplus \alpha_B\xi$, and $\eta\boxplus
\eta$. We described the \MWgrp{} $E(K)$ as a quotient of $H_2(B,\Z)$
by $T$, see eq.~\eqref{eq:MWses}. Therefore, the homology class of the
\MW{} sum $\mu \boxplus \nu$ is $\mu + \nu + (\text{something in }T)$.
That is
\begin{equation}
  \label{eq:plusboxplus}
  \mu\boxplus \nu = \mu + \nu -\sigma +
  (\text{linear combination of } F,\Theta_{1,1},\dots,\Theta_{3,2})
  \,,
\end{equation}
where we fixed the coefficient of $\sigma$ by the intersection number
with $F$ using the fact that $s\cdot F = 1$ for any section $s$. We
can compute the intersection number of any section with
$\mu\boxplus\nu$ from the height pairing and the structure of the
\MWgrp. The intersection with the remaining homology generators
$\Theta_{1,1},\dots,\Theta_{3,2}$ can simply be read off from
Figure~\ref{fig:sectionreducible}. Determining the coefficients in
eq.~\eqref{eq:plusboxplus} is then a linear algebra problem. We find
\begin{subequations}
\begin{align}
  \xi \boxplus \xi =&\, \xi + \xi - \sigma + \Theta_{2,1} +
  \Theta_{3,2}
  \\
  \alpha_B \xi \boxplus \xi =&\, \alpha_B \xi + \xi - \sigma -F +
  \Theta_{3,1} + \Theta_{3,2}
  \\
  \label{eq:etaplusxi}
  \eta \boxplus \xi =&\, \eta + \xi - \sigma -F + \Theta_{2,1} +
  \Theta_{3,1} + \Theta_{3,2}
  \\
  \eta \boxplus \alpha_B \xi =&\, \eta + \alpha_B\xi - \sigma -F +
  \Theta_{1,1} + \Theta_{1,2} + \Theta_{3,1}
  \\
  \eta \boxplus \eta =&\, \sigma + F -\frac{1}{3} \sum \Theta_{i,1}
  -\frac{2}{3} \sum \Theta_{i,2} = \notag \\ =&\, \eta + \eta - \sigma
  -F + \Theta_{1,1}+ \Theta_{2,1}+ \Theta_{3,1} \,.
\end{align}  
\end{subequations}

From this, we can determine the entire $G$ action on $H_2(B,\Z)$. The
result is summarized in Table~\ref{tab:homologyaction}.
\begin{table}[htbp]
  \renewcommand{\arraystretch}{1.4}
  \centering
  \begin{tabular}{c|c@{}c@{}c@{}c}
    $x$ & 
    $\displaystyle (\alpha_B)_\ast x$ & 
    $\displaystyle (t_\xi)_\ast x$ & 
    $\displaystyle (g_1)_\ast x = 
      \big( t_\xi \circ \alpha_B\big)_\ast x$  &
    $\displaystyle (t_\eta)_\ast x = (g_2)_\ast x$ 
    \\ \hline
    $\displaystyle \sigma$ & 
    $\displaystyle \sigma$ &
    $\displaystyle \xi$ &
    $\displaystyle \xi$ &
    $\displaystyle \eta$
    \\
    $\displaystyle F$ & 
    $\displaystyle F$ &
    $\displaystyle F$ &
    $\displaystyle F$ &
    $\displaystyle F$ 
    \\
    $\displaystyle \Theta_{1,1}$ & 
    $\displaystyle \Theta_{2,1}$ & 
    $\displaystyle \Theta_{1,1}$ & 
    $\displaystyle \Theta_{2,2}$ & 
    $\displaystyle \Theta_{1,2}$
    \\
    $\displaystyle \Theta_{2,1}$ & 
    $\displaystyle \Theta_{3,1}$ & 
    $\displaystyle \Theta_{2,2}$ & 
    $\displaystyle F-\Theta_{3,1}-\Theta_{3,2}$ & 
    $\displaystyle \Theta_{2,2}$
    \\
    $\displaystyle \Theta_{3,1}$ & 
    $\displaystyle \Theta_{1,1}$ & 
    $\displaystyle F - \Theta_{3,1} - \Theta_{3,2}$ & 
    $\displaystyle \Theta_{1,1}$ & 
    $\displaystyle \Theta_{3,2}$
    \\
    $\displaystyle \Theta_{1,2}$ & 
    $\displaystyle \Theta_{2,2}$ & 
    $\displaystyle \Theta_{1,2}$ & 
    $\displaystyle F - \Theta_{2,1} - \Theta_{2,2}$ & 
    $\displaystyle F - \Theta_{1,1} - \Theta_{1,2}$
    \\
    $\displaystyle \Theta_{2,2}$ & 
    $\displaystyle \Theta_{3,2}$ & 
    $\displaystyle F - \Theta_{2,1} - \Theta_{2,2}$ & 
    $\displaystyle \Theta_{3,1}$ & 
    $\displaystyle F - \Theta_{2,1} - \Theta_{2,2}$
    \\
    $\displaystyle \Theta_{3,2}$ & 
    $\displaystyle \Theta_{1,2}$ & 
    $\displaystyle \Theta_{3,1}$ & 
    $\displaystyle \Theta_{1,2}$ & 
    $\displaystyle F - \Theta_{3,1} - \Theta_{3,2}$
    \\[3mm]
    $\displaystyle \xi$ & 
    $\displaystyle \alpha_B\xi$ & 
    $\displaystyle\renewcommand{\arraystretch}{1} \begin{array}{c} 
      2 \xi - \sigma + 
      \hfill \\ \hfill
      {}+\Theta_{2,1}+\Theta_{3,2}
    \end{array}$ &
    \small
    $\displaystyle\renewcommand{\arraystretch}{1} \begin{array}{c} 
      \alpha_B \xi + \xi -\sigma-
      \hfill \\ \hfill
      {}-F+\Theta_{31}+\Theta_{32}
    \end{array}$ &
    \small
    $\displaystyle\renewcommand{\arraystretch}{1} \begin{array}{c} 
      \eta + \xi - \sigma -F +
      \hfill \\ \hfill
      {}+ \Theta_{2,1} + \Theta_{3,1} + \Theta_{3,2}
    \end{array}$
    \\[8mm]
    $\displaystyle \alpha_B \xi$ & 
    \small
    $\displaystyle\renewcommand{\arraystretch}{1} \begin{array}{c} 
      - \alpha_B\xi -\xi + 
      \hfill \\ \hfill
      {}+ 3\eta + \Theta_{1,1} + 
      \hfill \\ \hfill
      {}+\Theta_{2,1} + \Theta_{3,1}
    \end{array}$ &
    \small
    $\displaystyle\renewcommand{\arraystretch}{1} \begin{array}{c} 
      \alpha_B \xi + \xi-\sigma-F+
      \hfill \\ \hfill
      {}+\Theta_{3,1}+\Theta_{3,2}
    \end{array}$ &
    \small
    $\displaystyle\renewcommand{\arraystretch}{1} \begin{array}{c} 
      - \alpha_B\xi + 2\sigma + 2F -
      \hfill \\ \hfill
      {}- \Theta_{1,1} - \Theta_{1,2} -
      \hfill \\ \hfill
      {}- \Theta_{3,1}- \Theta_{3,2}
    \end{array}$ &
    \small
    $\displaystyle\renewcommand{\arraystretch}{1} \begin{array}{c} 
      \eta + \alpha_B\xi - \sigma -F +
      \hfill \\ \hfill
      {}+ \Theta_{1,1} + \Theta_{1,2} + \Theta_{3,1}
    \end{array}$
    \\[8mm]
    $\displaystyle \eta$ &
    $\displaystyle \eta$ &
    \small
    $\displaystyle\renewcommand{\arraystretch}{1} \begin{array}{c} 
      \eta + \xi - \sigma -F +
      \hfill \\ \hfill
      {}+ \Theta_{2,1} + \Theta_{3,1} + \Theta_{3,2}
    \end{array}$ &
    \small
    $\displaystyle\renewcommand{\arraystretch}{1} \begin{array}{c} 
      \eta + \xi - \sigma -
      \hfill \\ \hfill
      {}-F + \Theta_{2,1}+
      \hfill \\ \hfill
      {}+ \Theta_{3,1} + \Theta_{3,2}
    \end{array}$ &
    \small
    $\displaystyle\renewcommand{\arraystretch}{1} \begin{array}{c} 
      2\eta - \sigma -F +
      \hfill \\ \hfill
      {}+ \Theta_{1,1}+ \Theta_{2,1}+ \Theta_{3,1}
    \end{array}$
  \end{tabular}
  \caption{Summary of the $G$ action on $H_2(B,\Z)$.}
  \label{tab:homologyaction}
\end{table}
Instead of repeating the same arguments over and over again, we will
only discuss a few representative cases:
\begin{descriptionlist}
\item[$(\alpha_B)_\ast \sigma$:] By definition $\alpha_B$ fixes the
  $0$-section $\sigma$. Hence $(\alpha_B)_\ast\sigma = \sigma$.
\item[$(t_\xi)_\ast \sigma$:] Translating any section by $\xi$ is just
  the sum in the \MWgrp. This is then
  \begin{equation}
    (t_\xi)_\ast \sigma = \sigma \boxplus \xi = \xi
    \,.
  \end{equation}
\item[$(\alpha_B)_\ast \Theta_{1,1}$:] Since $\alpha_B$ cyclically
  permutes the $I_3$ singular fibers, the $\Theta_{1,1}$ component is
  just mapped to the $\Theta_{2,1}$ component: $(\alpha_B)_\ast
  \Theta_{1,1}=\Theta_{2,1}$.
\item[$(t_\xi)_\ast \Theta_{3,1}$:] The translation by $\xi$ rotates
  the $3$rd singular $I_3$ fiber according to
  Figure~\ref{fig:sectionreducible}. Therefore, it cyclically permutes
  the irreducible components
  $\Theta_{3,1}\to\Theta_{3,0}\to\Theta_{3,2}\to\Theta_{3,1}$. But
  $\Theta_{3,0}$ is not part of our chosen set of generators. We must
  eliminate it using the relation $F=\sum\Theta_{3,j}$. Hence,
  $(t_\xi)_\ast \Theta_{3,1} = \Theta_{3,0} =
  F-\Theta_{3,1}-\Theta_{3,2}$.
\end{descriptionlist}
and so on.

\subsection{Invariant Homology}
\label{sec:surfaceinvarianthomology}

Now that we have identified the $G_1=\big\{1,g_1,g_1^2\big\}$ and
$G_2=\big\{1,g_2,g_2^2\big\}$ actions on the homology lattice, it is a
simple exercise in linear algebra to find the invariant sublattice.
Picking a basis for $H_2(B,\Z)$, we can express $(g_1)_\ast$ and
$(g_2)_\ast$ as commuting $10\times 10$ matrices. The invariant
homology is then precisely the $+1$ eigenspace of these matrices,
which is straightforward to compute. Both of them are
$4$-dimensional, and a convenient choice of generators is
\begin{equation}
  H_2(B,\Z)^{G_1} = 
  \Span_\Z\Big\{ 
    F ,~
    -\sigma+\Theta_{2,1}+\eta ,~ 
    \Theta_{1,1} + \Theta_{3,1} +\Theta_{2,2} ,~
    \Theta_{3,1} + \Theta_{3,2} + 2 \xi + \alpha_B \xi
  \Big\}
\end{equation}
and
\begin{multline}
  H_2(B,\Z)^{G_2} = \Span_\Z\Big\{ F ,~
  \Theta_{1,1}+\Theta_{2,1}+\Theta_{3,1}+3 \eta ,~
  \\
  -\Theta_{1,1} + \Theta_{3,1} -\Theta_{2,2}+\Theta_{3,2} + 3 \xi + 3
  \alpha_B \xi ,~
  \\
  -\sigma + \Theta_{1,1} + \Theta_{2,1} + \Theta_{2,2} -\xi -2
  \alpha_B \xi + \eta \Big\} \,.
\end{multline}

We are, of course, interested in the $G=G_1\times G_2$ invariant
subspace. This is the intersection
\begin{equation}
  H_2(B,\Z)^G = 
  H_2(B,\Z)^{G_1} 
  \cap
  H_2(B,\Z)^{G_2} 
  \,.
\end{equation}
To prove this, we have to show that the inclusions ``$\subseteq$'' and
``$\supseteq$'' hold simultaneously. The first inclusion is true,
since a $G$ invariant homology class is necessarily invariant under
the subgroups $G_1$ and $G_2$. For the inclusion in the other
direction, note that every element $g\in G$ can be written as a
product
\begin{equation}
  g = g_1^{n_1} g_2^{n_2}
  ,\, \quad
  n_1, n_2=0,1,2
  \,.
\end{equation}
Therefore, a homology class preserved by $(g_1)_\ast$ and $(g_2)_\ast$
is also preserved by $g_\ast = (g_1^{n_1} g_2^{n_2})_\ast$. Hence, to
compute the $G$ invariant homology we only have to intersect the $G_1$
invariant subspace with the $G_2$ invariant subspace. This is again
simple linear algebra, and we only state the result. The
$G=\Z_3\times\Z_3$ invariant homology group has rank $2$ and is
generated by
\begin{equation}
  \label{eq:dP9invGhomologydef}
  t_1\eqdef F , \qquad t_2 \eqdef
  -\sigma+\Theta_{2,1}+\Theta_{3,1}+\Theta_{3,2} +2 \xi+ \alpha_B
  \xi+\eta - F
  \,.
\end{equation}
That is,
\begin{equation}
  \label{eq:dP9invGhomology}
  H_2(B,\Z) ^G = t_1 \Z \oplus t_2 \Z \,.
\end{equation}

\section{The \CY{} Threefold}
\label{sec:CY}

\subsection{The Fiber Product}
\label{sec:CYfiberproduct}

The fiber product of two $dP_9$ surfaces is a \CY{} threefold, as
already mentioned in Section~\ref{sec:CYintro}. We denote these two
$dP_9$ surfaces by $B_1$ and $B_2$ and the \CY{} threefold by $\Xt$.
Moreover, we want a $G=\Z_3\times\Z_3$ action $\Xt$ and, hence, on
$B_1$ and $B_2$ satisfying the constraints outlined in previous
sections. Therefore, we choose $B_1$ and $B_2$ to each correspond to
case 39a or one of the three special limits 39c, 68, and 68 in
Table~\ref{tab:singularfib}. By definition, these surfaces come with
projections
\begin{equation}
  \beta_j: B_j \to \CP^1
  ,\quad j=1,2
  \,.
\end{equation}
The fiber product $\Xt$ is defined as the hypersurface
\begin{equation}
  \Xt \eqdef
  B_1 \times_{\CP^1} B_2 \eqdef
  \Big\{ 
    (p_1,p_2) \in B_1\times B_2
  \Big|~
    \beta_1(p_1) = \beta_2(p_2)
  \Big\}
\end{equation}
within $B_1\times B_2$. The one equation in a $\dim_\C(B_1\times
B_2)=4$ dimensional space defines a $\dim_\C\big(\Xt\big)=3$
dimensional hypersurface, as desired. Furthermore, as was shown
in~\cite{dP9invariant}, the first Chern class is
\begin{equation}
  c_1\big(\Xt\big) = 0
  \,.
\end{equation}
Hence, $\Xt$ is a \CY{} threefold.

There is the following subtlety in this construction. If we consider
only one of the $dP_9$ surfaces, say $B_1$, then the choice of
coordinates on the base $\CP^1$ does not matter. That is, pick
$PGL(2,\C)\owns \tau: \CP^1\to \CP^1$. Then $\beta_1:B_1\to \CP^1$ and
$\tau \circ \beta_1:B_1\to \CP^1$ are two different projections. But,
since they are isomorphic, it makes little sense to distinguish them.
However, we are considering two $dP_9$ surfaces simultaneously.
Changing the coordinates on the two base $\CP^1$s relative to one
another does make a difference for the fiber product. This is clear
from the following description of the fiber product. The fiber product
$\Xt= B_1 \times_{\CP^1} B_2$ is a $T^4$ fibration over $\CP^1$, where
the fiber over $s\in \CP^1$ is precisely $\beta_1^{-1}(s) \times
\beta_2^{-1}(s)$. Changing $\beta_1$ to $\tau\circ\beta_1$ then
changes which fibers of $B_1$ and $B_2$ are paired up, so it changes
the fiber product.

Therefore, we must be careful with the relative choice of coordinates
implicit in the projections $\beta_1$ and $\beta_2$. To accomplish our
goal, we must choose the projections so that
\begin{itemize}
\item the $G$ action extends to the fiber product, that is, the
  hypersurface $\Xt\subset B_1\times B_2$ is preserved and
\item the $G$ action is free on $\Xt$, that is, the hypersurface is
  disjoint from the fixed point set in $B_1\times B_2$.
\end{itemize}

Let us first discuss under what conditions the $G$ action extends to
$\Xt$. Recall that we have two generators, $g_1$ and $g_2$, where
$g_1$ rotates the base $\CP^1$ while $g_2$ does not. Since $g_2$ keeps
every fiber stable, its action always extends to the fiber product. On
the other hand side, $g_1$ moves the fibers of $B_1$ and $B_2$.
Therefore, we must ensure that the fibers paired up in the fiber
product stay together under the $g_1$ action. That is, if
\begin{equation}
  F_j \subset B_j
  \,,\quad
  j=1,2
\end{equation}
are two fibers, then
\begin{equation}
  \beta_1(F_1) = \beta_2(F_2)
  \quad \Rightarrow \quad
  \beta_1\Big(g_1(F_1)\Big) = \beta_2\Big(g_1(F_2)\Big)  
  \,.
\end{equation}
This means that the induced action on the base $\CP^1$ must be the
same. In particular, the two fixed points on the base $\CP^1$ must be
the same. We take these two fixed points to be
\begin{equation}
  \label{eq:zeroinftyP1}
  [0:1],~[1:0] \in \CP^1
  \,.
\end{equation}
Henceforth, we require that the induced action on the base $\CP^1$ is
the same, that is 
\begin{equation}
  \beta_1 \circ g_1 \circ \beta_1^{-1} 
  = 
  \beta_2 \circ g_1 \circ \beta_2^{-1}
  \,.
\end{equation}
As we have seen, this implies that $\beta_j$ projects the two $g_1$
stable fibers of $B_j$ down to the two special points in
eq.~\eqref{eq:zeroinftyP1} for $j=1,2$.

Therefore, we have chosen the projections $\beta_1$ and $\beta_2$ so
that the $G$ action extends to $\Xt$. However, this does not fix the
projections uniquely. We still have both a continuous and a discrete
choice, which we will now use to obtain a free $G$ action. Recall that
$G$ acts freely on $\Xt$ if and only if for every subgroup of $G$
there are never two fibers containing fixed points which are paired in
the fiber product.  That is, for each fiber $\beta_1^{-1}(s) \times
\beta_2^{-1}(s)$ of $\Xt$ and for each $g\in G$, at most one of the
fibers $\beta_1^{-1}(s)$ of $B_1$ and $\beta_2^{-1}(s)$ of $B_2$
contains a point fixed by $g$.
\begin{figure}[htbp]
  \centering
  \input{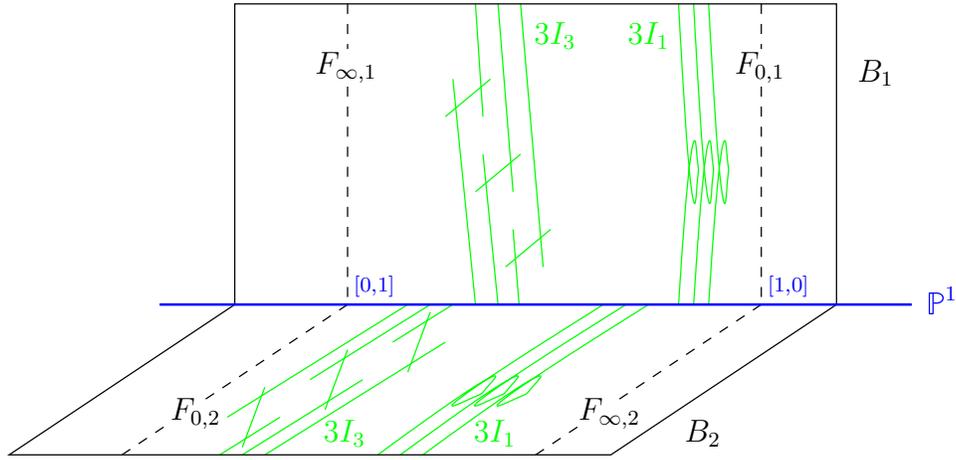}
  \caption{The fiber product $B_1\times_{\CP^1}B_2$, 
    showing our identifictaion of fibers.}
  \label{fig:fiberproduct}
\end{figure}
We have to distinguish the cases where $s$ is or is not one of the
special points eq.~\eqref{eq:zeroinftyP1}. First, assume that
$s\not=[0:1], [1:0]$. Furthermore, assume that $\beta_2^{-1}(s)$
contains a fixed point of $g\in G$. Then, we must show that we can
choose $\beta_1$ such that $\beta_1^{-1}(s)$ does not have a $g$ fixed
point. This can be achieved using the remaining continuous rescaling
\begin{equation}
  \label{eq:CP1rescale}
  \CP^1\to \CP^1,~
  [z_0:z_1] \mapsto 
  [\lambda z_0:z_1]
  \,,
  \quad 
  \lambda \in \C-\{0\}
  \,.
\end{equation}
Since there are only a finite number of special fibers
$\beta_2^{-1}(s)$, one can always find a suitable rescaling $\lambda$.
Hence, without loss of generality, we assume that $\beta_1$ and
$\beta_2$ are chosen so that the fiber product $\Xt$ does not have any
fixed points except, possibly, over $[0:1], [1:0]$. To exclude fixed
points in these two fibers, note that our $dP_9$ surfaces
$B_j$, $j=1,2$ were constructed so that for every $g\in G$
there is never a $g$ fixed point in both $\beta_j^{-1}([0:1])$ and
$\beta_j^{-1}([1:0])$ simultaneously. Hence, we can use the remaining
discrete choice in the $\CP^1$ coordinates to ensure that no fibers
with fixed points are paired up. This remaining choice is to exchange
the homogeneous coordinates on $\CP^1$, thus exchanging $[0:1]$ and
$[1:0]$. Doing this, if necessary, ensures that there are no fixed
points in $\Xt$ over $[0:1], [1:0]$ as well. The structure of such
a fiber product is illustrated in Figure~\ref{fig:fiberproduct}.

To summarize, one can always choose the projections $\beta_j:B_j\to
\CP^1$ in such a way that $G$ acts freely on the fiber product $\Xt$.
In the following, we will assume that this is the case. By
construction, this \CY{} threefold $\Xt$ is elliptically fibered with
respect to each of the two projections
\begin{equation}
  \pi_j: \Xt \to B_j
  \,, \quad j=1,2
  \,.
\end{equation}

\subsection{Homology of the Fiber Product}
\label{sec:homologyCYX}

We have constructed a specific family of \CY{} threefolds
\begin{equation}
  \Xt= B_1 \times_{\CP^1} B_2
  \,.
\end{equation}
Now, we want to determine its homology and Hodge numbers. 
We refer the reader to~\cite{dP9invariant} to explicit proofs. 

First, note that the fiber product is simply connected
\begin{equation}
  \pi_1\big(\Xt\big) = 1
  \,.
\end{equation}
Furthermore, $\Xt$ can be glued from $T^2$ bundles as follows. Recall
that $\Xt$ can be thought of as a $T^2\times T^2$ fibration over
$\CP^1$. Choose a sufficiently fine covering of $\CP^1$. Then,
locally, one of the two possible $T^2$ fibrations is actually smooth.
This implies the vanishing of the Euler characteristic
\begin{equation}
  \chi\big( \Xt \big) = 0
  \,.
\end{equation}
Moreover, it can be shown that the second cohomology group
$H^2\big(\Xt,\Z\big)\simeq \Pic\big(\Xt\big)$ is
\begin{equation}
  \label{eq:H2Xtquotient}
  H^2\Big(\Xt,\Z\Big) \simeq 
  \frac{ H^2(B_1,\Z) \oplus H^2(B_2,\Z) }{H^2(\CP^1,\Z)}
  \,,
\end{equation}
Counting dimensions, that is, ignoring torsion, and using
eq.~\eqref{eq:dP9cohomology} we immediately find that
\begin{equation}
  H^2\Big(\Xt,\Q\Big) = \Q^{19}
  \,.
\end{equation}
This determines the Hodge diamond as follows. The second Betti number
is 
\begin{equation}
  b_2 = h^{0,2} + h^{1,1} + h^{2,0} = 19
  \,.
\end{equation}
For a \CY{} threefold, $h^{0,2}$ and $h^{2,0}$ always vanish.
Therefore, $h^{1,1}=19$. Finally, the Euler characteristic of any
\CY{} threefold is $2(h^{1,1}-h^{2,1})$. In our case, the Euler
characteristic vanishes and, hence, $h^{1,1}=h^{2,1}$. To summarize,
the Hodge diamond of $\Xt \eqdef B_1 \times_{\CP^1} B_2$ is
\begin{equation}
  \vcenter{\xymatrix@!0@=7mm@ur{
    1 &  0 &  0 & 1 \\
    0 & 19 & 19 & 0 \\
    0 & 19 & 19 & 0 \\
    1 &  0 &  0 & 1 
  }}
  \,.
\end{equation}

\subsection{Homology of the \CY{} Manifold $\mathbf{X}$}
\label{sec:homologyCYXt}

Having constructed the simply connected \CYm{} $\Xt$ with a
$G=\Z_3\times\Z_3$ action, we now define
\begin{equation}
  X = \Xt / G
  \,.
\end{equation}
Because the $G$ action is free and analytic, this quotient is again a
smooth K\"ahler manifold. In fact, there is a general
theorem\footnote{This is not obvious, and only holds for proper \CY{}
  threefolds. Naively, one might fear that the first Chern class of
  the quotient is a torsion cohomology class.  Equivalently, the
  holomorphic $(3,0)$ form $\Omega$ might not be invariant. A proof of
  the theorem can be found in~\cite{BeauvilleNonAbel}.} that a fixed
point free quotient of a \CY{} threefold is again a \CY{} threefold.

The \CY{} threefold $X$ clearly has fundamental group
\begin{equation}
  \pi_1(X) = \Z_3 \times \Z_3
  \,.
\end{equation}
Moreover, its Euler characteristic is easily computed to be
\begin{equation}
  \chi\Big(\Xt/G\Big) = 
  \frac{1}{|G|} \chi\Big(\Xt\Big) =  \frac{1}{9} \cdot 0 = 0
  \,.
\end{equation}
The individual Betti and Hodge numbers can be found as follows. In
general, the rational cohomology groups on $X$ are the invariant
cohomology groups on $\Xt$. That is,
\begin{equation}
  H^\ast\big( X,\Q\big) 
  =
  H^\ast\big( \Xt,\Q\big)^G
  \,.
\end{equation}
It follows from eq.~\eqref{eq:H2Xtquotient} that
\begin{equation}
  H^2\big( \Xt,\Q)^G = 
  \left(\frac{ H^2(B_1,\Q) \oplus H^2(B_2,\Q) }
    {H^2(\CP^1,\Q)}\right)^G
  =
  \frac{ H^2(B_1,\Q)^G \oplus H^2(B_2,\Q)^G }{H^2(\CP^1,\Q)}
  \,.
\end{equation}
The invariant cohomology of each $dP_9$ surfaces is Poincar\'e dual to
the invariant homology given in eq.~\eqref{eq:dP9invGhomology}. By
counting dimensions, we compute the second Betti number of $X$ to be
$b_2=2+2-1=3$. By the same argument as in the previous section, this
determines all Hodge numbers of the \CY{} threefold $X$. The Hodge
diamond of $X=\Xt/G$, therefore, is
\begin{equation}
  \vcenter{\xymatrix@!0@=7mm@ur{
    1 &  0 &  0 & 1 \\
    0 &  3 &  3 & 0 \\
    0 &  3 &  3 & 0 \\
    1 &  0 &  0 & 1 
  }}
  \,.
\end{equation}

Note that there are $h^{2,1}=3$ complex structure moduli. These can be
understood as follows. Recall that each of the two surfaces $B_1$ and
$B_2$ come with a single parameter. This accounts for $2$ of the $3$
moduli. The third modulus is the relative scaling that is implicit in
the projections $\beta_j:B_j\to \CP^1$, see eq.~\eqref{eq:CP1rescale}.

\section*{Acknowledgments}

Volker Braun and Burt Ovrut are supported in part by the NSF Focused
Research Grant DMS~0139799 the DOE under contract
No.~DE-AC02-76-ER-03071. Tony Pantev is supported in part by NSF
grants DMS~0099715, DMS~0139799, and an A.~P.~Sloan Research
Fellowship. Ren\'e Reinbacher is supported by the Department of
Physics and Astronomy of Rutgers University under grant
No.~DOE-DE-FG02-96ER40959.

\appendix
\section{Limits of the Weierstrass Equation}
\label{sec:weierstrasslimit}

The resolution of the zero set of a Weierstrass equation
\begin{equation}
  y^2 z = x^3 + a(t) x z^2 + b(t) z^3
\end{equation}
is a rational elliptic surface. It is well-known how to read off
the singular fibers. The fibration degenerates whenever the
discriminant $D\eqdef 4 a(t)^3 + 27 b(t)^2$ vanishes. At each such
zero, the type of Kodaira fiber is determined by the order of
vanishing of the discriminant and the coefficients as in
Table~\ref{tab:weierstrassdegrees}.

For our one parameter Weierstrass equation~\eqref{eq:weierstrass}, the
singular fibers will change at special points of the parameter
$\gamma$. Hence, the \MWgrp{} also changes. We find the singular
fibers presented in Table~\ref{tab:Wdegenerations}.
\begin{table}[htbp]
  \centering
  \renewcommand{\arraystretch}{1.3}
  \begin{tabular}{c|c}
    Parameter & Singular fibers    
    \\ \hline
    generic &  
    $3I_0$,~$3I_3$ 
    \\
    $\gamma = -\frac{1}{24}$ & 
    $3I_0$,~$I_9$
    \\
    $\gamma = -\frac{5}{216}$ & 
    $4 I_3$
    \\
    $\gamma\to\infty$,~$t\gamma^{-\frac{1}{3}}$~fixed &
    $4 IV$
  \end{tabular}
  \caption{Singular fibers of the $1$-parameter family 39a and 
    limits thereof.}
  \label{tab:Wdegenerations}
\end{table}

\section{Orbifold Resolutions}
\label{sec:resolutions}

Let $G$ be an arbitrary finite group acting on a rational elliptic
surface $B\to \CP^1$. Then the quotient $B/G$ has orbifold
singularities. We want to show that the minimal resolution
$\widehat{B/G}$ is again a rational elliptic surface. This relies on
the classification of algebraic surfaces. We review here some
fundamental facts.

The plurigenera $P_n(X)$ of any surface $X$ are the number of
sections of a certain line bundle,
\begin{equation}
  P_n(X) \eqdef \dim H^0\Big( \Osheaf_X[nK] \Big)
  \,.
\end{equation}
Their asymptotic growth with $n \in \Z$, $n\geq 0$ is the most
important birational invariant of the surface $X$. It is called the
Kodaira dimension and takes values in
\begin{equation}
  \kappa(X) \in \{ -\infty, 0, 1, 2 \}
\end{equation}
By definition of $\kappa$, the plurigenera $P_n$ grow like $n^\kappa$.
What is the Kodaira dimension of a rational elliptic surface $B$?
Recall that $B$ is the blow up of $\CP^2$ at $9$ points. Since the
Kodaira dimension is a birational invariant, we may just as well
compute the Kodaira dimension of $\CP^2$. But all plurigenera
\begin{equation}
  P_n\big(\CP^2\big) = 0
  \,.
\end{equation}
Hence, the Kodaira dimensions are
\begin{equation}
  \kappa(B) = \kappa\big(\CP^2\big) = -\infty
  \,.
\end{equation}

Now, the $G$ action preserves the line bundle $\Osheaf_B[nK]$ and,
therefore, defines an action on $H^0\big( \Osheaf_B[nK] \big)$. The
plurigenera of the quotient are the dimensions of the invariant parts
\begin{equation}
  P_n\big(\widehat{B/G}\big) 
  = 
  \dim H^0\Big(  \Osheaf_B[nK] \Big)^G
  \leq 
  \dim H^0\Big(  \Osheaf_B[nK] \Big)
  = 0
  \,.
\end{equation}
Therefore,
\begin{equation}
  \kappa\big( \widehat{B/G} \big) = -\infty
  \,.
\end{equation}
Another birational invariant is the irregularity $q$. In our case,
\begin{equation}
  q\big( \widehat{B/G} ) = 
  \dim H^0\big(\Omega^1\big)^G = 0 =
  \dim H^0\big(\Omega^1\big) = q(\CP^2) = q(B)
  \,.
\end{equation}
By the classification of algebraic surfaces, a surface of Kodaira
dimension $\kappa=-\infty$ and irregularity $q=0$ is rational.
Furthermore, the $G$ action has to preserve the elliptic fibration of
$B$, as proven in Section~\ref{sec:CYintro}. Hence, the quotient $B/G$
and its resolution $\widehat{B/G}$ are also elliptically fibered.
Therefore, $\widehat{B/G}$ is again a rational elliptic surface.

\section{A Synthetic Approach}
\label{sec:synthetic}

While we argued that one can read off the intersection pattern in
Figure~\ref{fig:sectionreducible} from the Weierstrass equation, it
can actually be determined on general grounds without having to refer
to any geometric realization. We need only know that
\begin{itemize}
\item The rational elliptic surface $B$ has $3I_3$ and $3I_1$ singular
  fibers.
\item There exists an order $3$ section $\eta \in E(K)$, yielding a
  $\Z_3$ action $t_\eta: B\to B$.
\item There exists another $\Z_3$ action $\alpha_B$ on $B$, fixing the
  $0$-section and acting non-trivially on the base $\CP^1$.
\item $\alpha_B$ has isolated fixed points on $\beta^{-1}([0,1])$ and,
  in addition, fixes $\beta^{-1}([1,0])$ point-wise.
\item $t_\eta$ and $\alpha_B$ commute.
\end{itemize}

We know the action of $\alpha_B$ on all homology generators except on
the free part of the \MWgrp. Now, the Euler characteristic of the
$\alpha_B$ fixed point set is $\chi(B^{\alpha_B})=3$. The Lefschetz
fixed point formula then determines the entire $\alpha_B$ action. The
result is that $(\alpha_B)_\ast$ rotates $E(K)_{\mathrm{free}} =
A_2^\ast$ by $120^0$. We can now pick any section $\xi$ and define
$g_1\eqdef t_\xi\circ \alpha_B$. Together with $g_2\eqdef t_\eta$ this
generates a $\Z_3\times \Z_3$ group action on $B$. All $g_1$ fixed
points are contained in $\beta^{-1}([0,1])$ if only $\xi_{[1,0]}\not=
0$. We ensure this by choosing a section $\xi$ which does not
intersect the $0$-section, $\xi \sigma=0$. The section $\xi$ is then a
minimal length point of the $A_2^\ast$ lattice. Now, $g_2$ necessarily
fixes the singular point on the $I_1$ fibers. Therefore, the quotient
has a singularity modeled on $\C^2/\Z_3$. Resolving this contributes
$3\cdot 2=6$ to the Euler characteristic, so the only way to get
$\chi(\widehat{B/G_2})=12$ is if $g_2$ acts freely on every other
fiber. To summarize: $g_1$ fixes $3$ points in $\beta^{-1}([0,1])$ and
$g_2$ fixes one point in each of the three $I_1$ singularities.

Now, assume that $\eta$ intersects the neutral component of the
$i^\text{th}$ $I_3$ fiber. Then, there must be at least $3$ fixed
points in this fiber, but there are actually none. Therefore, $\eta$
cannot intersect $\Theta_{i,0}$. We choose our notation so that $\eta$
intersects $\Theta_{i,1}$.

Finally, using the height pairing, we obtain
\begin{equation}
  \frac{2}{3} = 
  \left\langle \xi, \xi \right\rangle =
  2 - \sum_s \contr_s(\xi,\xi)
  \,.
\end{equation}
Since $\contr_s(\xi,\xi)$ is either $0$ or $\frac{2}{3}$, we see that
$\xi$ has to intersect precisely one of the $I_3$ fibers in the
neutral component, which we call $\Theta_{1,0}$. By cyclic symmetry
(that is, the $\alpha_B$ action), $\xi$ must intersect the other two
$I_3$ fibers such that
\begin{itemize}
\item in one $I_3$ fiber, $\xi$ and $\eta$ intersect the same
  irreducible component and
\item in the last $I_3$ fiber, $\xi$ and $\eta$ intersect different
  irreducible components.
\end{itemize}
We fix our notation completely by asking that $\xi$ intersects
$\Theta_{2,1}$ and $\Theta_{3,2}$. To summarize, we obtained precisely
the intersection pattern in Figure~\ref{fig:sectionreducible}.

\section{$\alpha_B$-Invariant Homology}
\label{sec:alphahomology}

Although we do not need the $\alpha_B$ invariant part of the homology
group in this paper, we record it here for future use. We
already know the action of $\alpha_B$ on the homology, so it is again
straightforward to identify the $+1$ eigenspace. The invariant
sublattice has rank $4$ and is generated by
\begin{equation}
  H_2(B,\Z)^{\alpha_B} = 
  \Span\!\Big( 
  \sigma,~
  F,~
  \eta,~
  \sum_{i=1}^3 \Theta_{i,1}
  \Big)
  \,.
\end{equation}

\bibliographystyle{JHEP} \renewcommand{\refname}{Bibliography}
\addcontentsline{toc}{section}{Bibliography} \bibliography{main}

\end{document}